\documentclass[aps,preprint,showpacs,superscriptaddress,groupedaddress]{revtex4}  

\usepackage{graphicx}  
\usepackage{float}
\usepackage{dcolumn}   
\usepackage{bm}        
\usepackage{amssymb}   
\usepackage{slashed}   
\usepackage{amsmath}   
\usepackage{simplewick} 
\usepackage{verbatim}  
\usepackage{color} 
\usepackage{appendix} 
\usepackage{epstopdf}
\usepackage{subfig}
\usepackage[percent]{overpic}

\usepackage[bookmarks=true,colorlinks=true,linkcolor=blue,unicode=true]{hyperref}

\newcommand{\beginsupplement}{%
        \setcounter{table}{0}
        \renewcommand{\thetable}{A\arabic{table}}%
        \setcounter{figure}{0}
        \renewcommand{\thefigure}{A\arabic{figure}}%
        \setcounter{equation}{0}
        \renewcommand{\theequation}{A\arabic{equation}}%
     }

\begin{document}

\widetext

\title{Shape of a skyrmion}

\author{Ji-Chong Yang}
\affiliation{Department of Physics  \&  State Key Laboratory of Surface Physics,   Fudan University,\\ Shanghai 200433, China}

\author{Qing-Qing Mao}
\affiliation{Department of Physics  \&  State Key Laboratory of Surface Physics,   Fudan University,\\ Shanghai 200433, China}

\author{Yu Shi}
\affiliation{Department of Physics  \&  State Key Laboratory of Surface Physics,   Fudan University,\\ Shanghai 200433, China}

\affiliation{Collaborative Innovation Center of Advanced Microstructures, Fudan University, \\Shanghai 200433, China}

%
%
%
\vskip 0.25cm

\date{\today}

\begin{abstract}
We propose a method of determining the shape of a two-dimensional magnetic skyrmion, which can be parameterized as the position dependence of the orientation of the local magnetic moment,  by using the expansion in terms of the eigenfunctions of the Schr\"{o}dinger equation of a harmonic oscillator. A variational calculation is done, up to  the next-to-next-to-leading order. This result is verified by  a lattice simulation based on Landau-Lifshitz-Gilbert equation. Our method is also applied to the dissipative matrix in the Thiele equation as well as  two interacting skyrmions in a bilayer system.
\end{abstract}

\maketitle

\section{\label{sec:1}Introduction}

A magnetic skyrmion is a two-dimensional~(2D) topologically protected swirling spin texture in a chiral magnet~\cite{nagaosa,Fert}. It was discovered in an intermetallic compound  MnSi by using neutron scattering~\cite{Muhlbauer}, and was also observed by using Lorentz transmission electron microscopy~(LTEM)~\cite{Yu1}, spin-resolved scanning tunnelling microscopy ~\cite{2DSkyrmion2}, etc. The critical current density for the manipulation of a magnetic skyrmion is much lower than that for magnetic domain walls~\cite{ultralow}. Therefore it was proposed that skyrmions  may act as future information carriers in magnetic information storage and processing devices.

For a typical 2D skyrmion, The orientation ${\bf n}$ of the local magnetic moment can be parameterized in cylindrical coordinates as~\cite{NParameterAndLinear,NParameterArcTan}
\begin{equation}
\begin{split}
&{\bf n}(\rho,\phi,z)=\sin [\theta (\rho)] {\bf e}_{\phi}+\cos [\theta (\rho)] {\bf e}_z,\\
\end{split}
\label{eq.1.1}
\end{equation}
where $\theta$ is the angle between ${\bf n}$ and ${\bf e}_z$.

$\theta (\rho)$ describes the shape of a skyrmion. On large scales, a skyrmion can be considered as a point-like particle because of its topological nature. However, on scales compatible with or smaller than its radius, the shape of a skyrmion should be taken into account and becomes an interesting subject. For example, the current-driven motion of a skyrmion can be described in terms of the Thiele equation~\cite{Thiele},  which is widely used  in studying  rotational property~\cite{ThieleStudy,ThieleStudyAndJDB}, skyrmion Hall effect~\cite{Jiang}, skyrmions in bilayer systems~\cite{bilayer1,bilayer2,Zhang2}, etc. The dissipative matrix in the Thiele equation relies on $\theta (\rho)$. Besides, $\theta (\rho)$ is important in studying the interaction between two skyrmions in bilayer systems~\cite{bilayer2}, because the interaction is significant only  when the two skyrmions are close to each other.

Hence it is important to determine  $\theta (\rho)$. Previously, $\theta (\rho)$ was obtained by numerical methods~\cite{pin} or was assumed to either be linear in $\rho$~\cite{Jiang,NParameterAndLinear,thetarho}, or satisfy $\theta(\rho) =\arctan \left(\exp\left(-\rho/\Delta\right)\right)$~\cite{NParameterArcTan,arctan}.
In this paper, instead, we propose an efficient method to analytically calculate $\theta (\rho)$ of a 2D skyrmion stabilized by an external magnetic field. We find an analytical expression of $\theta (\rho)$ and compare it with our numerical result. We also study the radii of the skyrmions by a lattice simulation of the Landau-Lifshitz-Gilbert~(LLG) equation. Using the  approximation in our analytical method, the dissipative matrix element and the interaction between the skyrmions can be explicitly expressed in terms of the ferromagnetic coupling, the   Dzyaloshinskii-Moriya~(DM) interaction strength and the magnetic field. Our analytical result is confirmed by our numerical result.

The rest of the paper is organized as the following. Our analytical method is introduced in Sec.~\ref{sec:2}. Our lattice simulation is discussed   in Sec.~\ref{sec:3}. The dissipative matrix element and the interaction between two skyrmions are discussed in Sec.~\ref{sec:4}. A summary is made in Sec.~\ref{sec:5}.

\section{\label{sec:2}Harmonic Oscillator Expansion}

First we briefly review the equation of $\theta(\rho)$ in Eq.~(\ref{eq.1.1}), as well as the numerical solution following the method used in~\cite{pin}.

The Hamiltonian can be written in terms of the dimensionless parameters as~\cite{FreeEnergy}
\begin{equation}
\begin{split}
&{\mathcal H}_{\rm tot}({\bf r})=\frac{J}{2}\nabla {\bf n}\cdot \nabla {\bf n} +D {\bf n}\cdot (\nabla \times {\bf n})-{\bf B}\cdot {\bf n},\\
\end{split}
\label{eq.2.1}
\end{equation}
where ${\bf n}$ is the orientation of the local magnetic moment,
given above in Eq.~(\ref{eq.1.1}), $J$ is the local ferromagnetic exchange strength, $D$ is the local strength of DM interaction, ${\bf B}$ is the  external magnetic field. $\theta (\rho)$ can be obtained by minimizing the total energy
\begin{equation}
\begin{split}
&F=\int d{\bf r} \mathcal{F}(\bf r)=2\pi \int d\rho \rho \mathcal{F}(\rho),\\
\end{split}
\label{eq.2.2}
\end{equation}
where the energy density $\mathcal{F}$ is~\cite{pin}
\begin{equation}
\begin{split}
&\mathcal{F}(\rho)={\mathcal H}_{\rm tot}(\rho)-{\mathcal H}_{\rm fe}=2J\left[\left(\frac{1}{2}\frac{\partial \theta}{\partial \rho}+\kappa\right)^2-\kappa^2+\frac{\sin^2(\theta)}{4\rho^2}+\frac{\kappa\sin(2\theta)}{2\rho}\right]-B(\cos (\theta)-1),\\
\end{split}
\label{eq.2.3}
\end{equation}
with $\kappa \equiv D/2J$, ${\mathcal H}_{\rm fe}=-B$. It has been assumed that the magnetic field  ${\bf B}=B({\bf r}){\bf e}_z$ is along $z$ direction. The Euler-Lagrange equation yields
\begin{equation}
\begin{split}
&\frac{ \sin(\theta)\cos(\theta)}{\rho}-\theta ' - \rho \theta ''-2\frac{D}{J}\sin^2(\theta)+\frac{B}{J} \rho \sin (\theta)=0,
\end{split}
\label{eq.2.4}
\end{equation}
which can be solved numerically by using the scheme of finite differences~\cite{pin}. In the rest of this paper, the numerical solutions are always  obtained by using this method.

The main content of our paper is the following  analytical method. A  function   $f(x)$ well defined in $[0,\infty)$, with $f(0)>0$, $f(\infty)=0$, and $\int _0^{\infty}dx |f(x)|^2$ being finite, can be expanded in terms of the eigenfunctions of the quantum Hamiltonian of an harmonic oscillator,
\begin{equation}
\begin{split}
&\phi_{n,\omega}(x)=\left(\frac{\omega}{\pi}\right)^{\frac{1}{4}}\frac{1}{\sqrt{2^n n!}}H_n(\sqrt{\omega}x)e^{-\frac{\omega x^2}{2}},\\
\end{split}
\label{eq.2.6}
\end{equation}
where $H_n(x)$ is the Hermite polynomials. $\phi_{n,\omega}(x)$'s are solutions of
\begin{equation}
\begin{split}
&\left(\frac{1}{2}\omega^2  x^2+\frac{1}{2}\frac{d^2}{dx^2}\right)\phi= E\phi.
\end{split}
\label{eq.2.5}
\end{equation}
Therefore   $f(x)$ can be expanded as
\begin{equation}
\begin{split}
&f(x)= \sum _{n=0}^{\infty}C_n\phi _{2n,\omega}(x)\\
\end{split}
\label{eq.2.7}
\end{equation}
with $C_n=\int _0^{\infty}dx f(x)\phi _{2n,\omega}(x)$.

Previous ansatz functions for $\theta (\rho)$ included the linear functions~\cite{Jiang,NParameterAndLinear,NParameterArcTan,thetarho}, as well as $ \tan^{-1}(\exp(-\rho/\Delta ))$~\cite{arctan,NParameterArcTan}. In  Appendix A, these ansatz functions are expanded in terms of harmonic oscillator functions, as examples.

\subsection{\label{sec:2.1}Leading order~(LO) approximation}

As the numerical solution is close to a Gaussian function, we assume that the LO wave function, as a Gaussian function, is a good approximation. Thus the coefficients in  the expansion $\theta(\rho)=\sum_n  C_n\phi _{2n,\omega}(\rho)$ satisfy $C_0\gg C_{i>0}$. Under this assumption, one can use the Rayleigh-Ritz variational method~\cite{RRMethod} to obtain  $\omega$ and $C_n$'s order by order.

At LO, $\theta(\rho)\approx \theta_{\rm LO}(\rho)=C_0\left(\omega/\pi\right)^{1/4}\exp (-\omega  \rho^2/2)$. To ensure $\theta (0)=\pi$ and $\theta (\infty)=0$, we use $\theta_{\rm LO} (\rho)=\pi \exp (-\omega  \rho^2/2)$ as a trial solution. To minimize the total energy $F$, $\omega $ should satisfy   $dF/d\omega = 0$,  leading to
\begin{equation}
\begin{split}
&\frac{D\pi^{\frac{3}{2}}}{2\omega\sqrt{2\omega}}-\frac{D}{4\omega}\int _0^{\infty}d\rho \sin (2\pi e^{-\frac{\omega \rho^2}{2}})-\frac{B}{2\omega^2}\left(2 (-\text{Ci}(\pi )+\gamma _E+\log (\pi ))\right)=0,
\end{split}
\label{eq.2.9}
\end{equation}
where $\gamma _E$ is the Euler constant, ${\rm Ci}$ is the cosine integral function  defined as
\begin{equation}
\begin{split}
&{\rm Ci}(z)\equiv -\int _z^{\infty}dt\frac{\cos (t)}{t}.
\end{split}
\label{eq.2.10}
\end{equation}
The solution can be obtained as
\begin{equation}
\begin{split}
&\omega_{\rm LO}=\left(\frac{\sqrt{2}B\left(2 (-\text{Ci}(\pi )+\gamma +\log (\pi ))\right)}{D(\pi^{\frac{3}{2}}-a_0)}\right)^2,\\
\end{split}
\label{eq.2.11}
\end{equation}
where
\begin{equation}
\begin{split}
&a_0\equiv\frac{1}{\sqrt{2}}\int _0^{\infty}dx \sin (2\pi e^{-\frac{x^2}{2}})\approx 0.250432,\\
\end{split}
\label{eq.2.12}
\end{equation}
\begin{equation}
\begin{split}
&\omega_{\rm LO}\approx 0.768548\left(\frac{B}{D}\right)^2.\\
\end{split}
\label{eq.2.13}
\end{equation}

\begin{figure}
\includegraphics[width=0.95\textwidth]{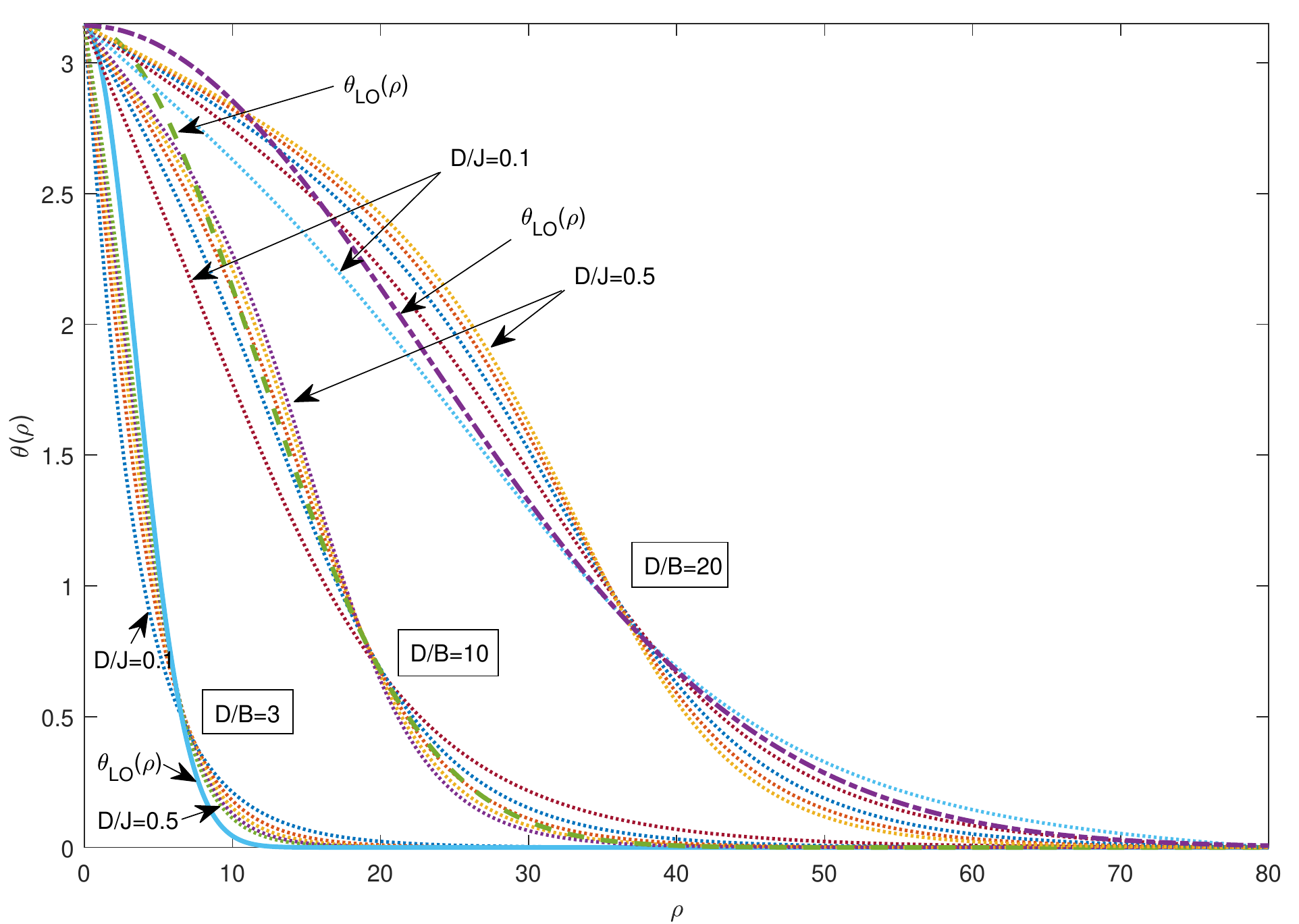}
\caption{\label{fig:3} The numerical result of $\theta (\rho)$ and  the function $\theta_{\rm LO}(\rho)$.  The solid line represents $\theta _{\rm LO}(\rho)$ for $D/B=3$, the dashed line in the middle represents $\theta _{\rm LO}(\rho)$ for $D/B=10$, while the dotted dashed line on the right represents $\theta _{\rm LO}(\rho)$ for $D/B=20$. Around each line of $\theta _{\rm LO}(\rho)$, the five dotted lines  represent the numerical results for the corresponding value of $D/B$ and for $D/J=0.1,0.2,0.3,0.4,0.5$. For each group  of five dotted lines with a same value of $D/B$, the slope increases   with $D/J$. }
\end{figure}

To verify the leading order approximation, some examples of the results from $\theta _{\rm LO}(\rho)$ together with the numerical results are shown in Fig.~\ref{fig:3}. The parameter values $D/J=0.2$ and $D/B=10$ or $20$ are close to $D/J=0.18$ and $B/J=0.0075$ to $0.0252$ in Ref.~\cite{ThieleStudyAndJDB}. The parameter values $D/J=0.1$ and $D/B=20$ are close to $D/J=0.09$ and $B/J=0.001875$ to $0.0063$ in Ref.~\cite{JDBRef1}. The parameter values $D/J=0.5$ and $D/B=3$ are close to $D/J=0.5$ and $B/J=0.15-0.3$ in Ref.~\cite{JDBRef2}.

We also find numerically that for a given value of $D/J$, the size of the skyrmion is nearly independent of $J$, and depends on $D/B$. This is in consistency with the LO approximation, in which $\theta _{\rm LO}(\rho)$ is Gaussian function, thus the radius of a skyrmion is $R\propto \left(\omega_{\rm LO}\right)^{-1/2} \propto D/B$, which is independent of $J$. This is also in consistency with the  previous results using dimensionless parameters~\cite{ThieleStudyAndJDB,JDBRef1,alpha3,bilayer2,pin,Tchoe}. Note that $J$ is merely an energy unit, and that $\rho$ is also dimensionless. For realistic materials, the radii of the skyrmions depend on realistic ferromagnetic coupling because the unit of $\rho$ depends on it, and the rescaling of $\rho$ must be taken into account. The relation between the dimensionless parameters and the realistic parameters will be discussed in Sec.~\ref{sec:3}.

\subsection{\label{sec:2.2}Next-to-next-to-leading order~(NNLO) approximation}

At the next-to-leading order~(NLO), the function $\theta (\rho)$   can be approximated as $\theta_{\rm NLO}(\rho)=\pi \exp\left(-\omega \rho^2/2\right)+c \pi \omega \rho ^2 \exp\left(-\omega \rho^2/2\right)$, where $c$ is a parameter. Solving the equations $dF/d\omega = 0$ and $dF/dc=0$, we find that the result is the same as the LO, that is, $c\approx 0$. So we need to consider the NNLO approximation. The trial solution at NNLO can be written as $\theta_{\rm NNLO}(\rho)=C_0 \phi _{0,\omega}+C_1 \phi _{2,\omega}+C_2 \phi _{4,\omega}$. It is  required that  $\theta_{\rm NNLO}(0)=\pi$ and $\theta_{\rm NNLO}(\infty)=0$, thus  $\theta_{\rm NNL}(\rho)$ can be parameterized as
\begin{equation}
\begin{split}
&\theta_{\rm NNLO}(\rho)=\pi e^{-\frac{\omega \rho^2}{2}}+a \pi \omega \rho ^2 e^{-\frac{\omega \rho^2}{2}}+b \pi  (\omega \rho ^2)^2 e^{-\frac{\omega \rho^2}{2}},\\
\end{split}
\label{eq.2.14}
\end{equation}
where $a$, $b$ and $\omega$ are parameters to be determined. From $dF/db=0$, $dF/da=0$ and $dF/d\omega=0$, one obtains
\begin{equation}
\begin{split}
&\frac{J}{2}\left(a_{i11}+a_{i12}a+a_{i13}b\right)+\frac{D}{2\sqrt{\omega}}\left(a_{i21}+a_{i22}a+a_{i23}b\right)\\
&-\frac{B}{2\omega}\left(a_{i31}+a_{i32}a+a_{i33}b\right)+\mathcal{O}(a^2)+\mathcal{O}(b^2)+\mathcal{O}(ab)= 0,\\
\end{split}
\label{eq.2.15}
\end{equation}
where $i=1,2,3$. The coefficients $a_{ijk}$ are constant numbers independent of $J$, $B$ and $D$, and $\theta_{\rm NNLO}(\rho)$ can be   expressed  in terms of these constant numbers and $J$, $B$ and $D$. The detailed  calculation is given  in the Appendix A. It is found that
\begin{equation}
\begin{split}
&a\approx 0.2021 +\frac{\frac{B}{J}(-0.4364\frac{B}{J}+0.1449\left(\frac{D}{J}\right)^2)}{\left(\frac{B}{J}\right)^2+0.01388\frac{B}{J}\left(\frac{D}{J}\right)^2-0.1148\left(\frac{D}{J}\right)^4},\\
&b\approx -0.09900 +\frac{\frac{B}{J}(0.2026\frac{B}{J}-0.06728\left(\frac{D}{J}\right)^2)}{\left(\frac{B}{J}\right)^2+0.01388\frac{B}{J}\left(\frac{D}{J}\right)^2-0.1148\left(\frac{D}{J}\right)^4},\\
&\omega_{\rm NNLO}\approx \left(\frac{B}{D}\frac{0.9594\left(\frac{B}{J}\right)^2-0.02628\frac{B}{J}\left(\frac{D}{J}\right)^2-0.09704\left(\frac{D}{J}\right)^4}{\left(\frac{B}{J}\right)^2+0.04050\frac{B}{J}\left(\frac{D}{J}\right)^2-0.1237\left(\frac{D}{J}\right)^4}\right)^2.\\
\end{split}
\label{eq.2.21}
\end{equation}

\begin{figure}
\includegraphics[width=0.95\textwidth]{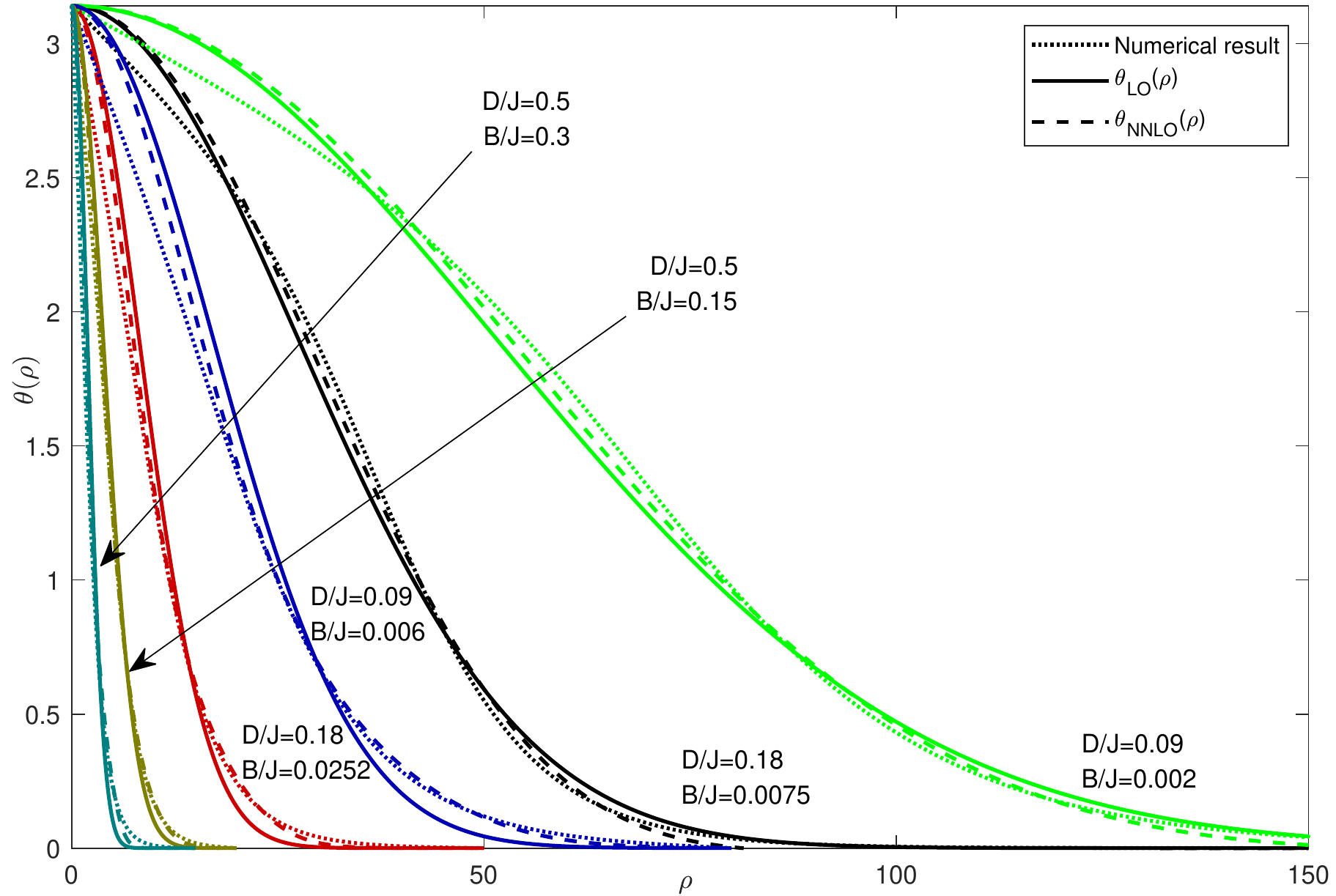}
\caption{The numerical result together with the results from  $\theta_{\rm LO}(\rho)$ and $\theta_{\rm NNLO}(\rho)$. The dotted lines represent numerical results, the solid lines represent $\theta _{\rm LO}$, the dashed lines represent $\theta _{\rm NNLO}$.  The parameters $J$, $D$ and $B$ are chosen to be such that  the  system is in the skyrmion phase, that is, $D/J=0.18$ while  $0.0075<B/J<0.0252$ \cite{ThieleStudyAndJDB},  $D/J=0.09$ while $0.001875<B/J<0.0063$\cite{JDBRef1}, and $D/J=0.5$ while  $0.15<B/J<0.30$ \cite{JDBRef2}.}
\label{fig:41}
\end{figure}

In Fig.~\ref{fig:41}, $\theta_{\rm LO}(\rho)$ and $\theta_{\rm NNLO}(\rho)$ are compared with the numerical results. The parameters $J$, $D$ and $B$ are chosen within the regimes of skyrmion phase,  with  $D/J=0.18$ while  $0.0075<B/J<0.0252$, $D/J=0.09$ while $0.001875<B/J<0.0063$, and  $D/J=0.5$ while $0.15<B/J<0.30$\cite{ThieleStudyAndJDB,JDBRef1,JDBRef2}. As expected, $\theta _{\rm NNLO}(\rho)$ is closer to the numerical results.

\section{\label{sec:3}Lattice Simulation}

To verify the  the variational calculation, we also study the radii of the skyrmions by doing the lattice simulation, which is  based on the LLG equation~\cite{nagaosa,pin,LLG,LLG2}
\begin{equation}
\begin{split}
&\frac{d}{dt} {\bf n}_{\bf r}=-{\bf B}_{\rm eff}({\bf r})\times {\bf n}_{\bf r}-\alpha {\bf n}_{\bf r}\times \frac{d}{dt} {\bf n}_{\bf r},\\
\end{split}
\label{eq.4.1}
\end{equation}
where ${\bf n}_{\bf r}$ is the local magnetic momentum at site ${\bf r}$, $\alpha$ is the Gilbert damping constant,
\begin{equation}
\begin{split}
&{\bf B}_{\rm eff}({\bf r})=-\frac{\delta H}{\delta {\bf n}_{\bf r}}
\end{split}
\label{eq.4.2}
\end{equation}
is the effective magnetic field, with the discrete Hamiltonian $H$~\cite{JDBRef1,ThieleStudyAndJDB}
\begin{equation}
\begin{split}
&H=\sum _{{\bf r},i=x,y}\left[-J({\bf r}){\bf n}_{{\bf r}+\delta_i}-D({\bf r}) {\bf n}_{{\bf r}+\delta_i}\times {{\bf e}}_i-{\bf B}\right]\cdot {\bf n}_{\bf r},\\
\end{split}
\label{eq.4.3}
\end{equation}
where $\delta _{i}$ refers to each neighbour, and $\delta _{i}= {\bf e}_{i}$ on a square lattice.  So~\cite{pin}
\begin{equation}
\begin{split}
&{\bf B}_{\rm eff}({\bf r})=\sum _{i=x,y}\left[J({\bf r}){\bf n}_{{\bf r}+\delta_i}+J({\bf r}-\delta_i){\bf n}_{{\bf r}-\delta_i}\right]\\
&+\sum _{i=x,y}\left[D({\bf r}){\bf n}_{{\bf r}+\delta_i}\times {{\bf e}}_i-D({\bf r}-\delta _i){\bf n}_{{\bf r}-\delta_i}\times {{\bf e}}_i\right]+{\bf B}({\bf r}).
\end{split}
\label{eq.4.4}
\end{equation}

We run the simulation on a $512\times 512$ square lattice. The simulation was  done by using the GPU~\cite{program}. The LLG is numerically integrated by using the fourth-order Runge-Kutta method. We run the simulation with $D/J=0.18$ and with $B/J=0.012,0.015,0.018,0.024$ such that $0.009<B/J<0.0252$, and it is set that $\alpha=0.04$~\cite{ThieleStudyAndJDB} and time step $\Delta t= 0.01$. To study the radii of skyrmions with different $J$, we run the simulation with $J=1,2,3,4,5$.

We study both  an isolated skyrmion and   the skyrmion phase. For the isolated skyrmion, we run the simulation with periodic boundary condition and with the initial state ${\bf n}_r={\bf e}_z$, except  ${\bf n}_r=-{\bf e}_z$ when the sites are near  the center, with $|{\bf r}-(255,255)|<10$. Under such initial condition, a single skyrmion can be created at the center,  due to the mechanism similar to that in Ref.~\cite{Romming}, where a skyrmion can be created at a desired position by flipping the spin at the position. To study the skyrmion phase, we run the simulation with open boundary condition and with randomized initial configurations.

To evaluate the radius $r_s$ of the skyrmion, one needs to calculate the number of the sites in the isoheight contour of $n_z=1$. However, $n_z$ only approaches $1$, so we use the isoheight contour of $n_z=0.5$ and compare $r_s$ with the solutions $\rho$ of the equations $\theta _{\rm LO}(\rho)=\cos^{-1}(0.5)$ and $\theta _{\rm NNLO}(\rho)=\cos^{-1}(0.5)$. The details on how the radii are obtained  can be found in Appendix, and the results are shown in Fig.~\ref{fig:rs}. One can find that the value of $J$ has little effect when  dimensionless parameters are used, as discussed in Sec.~\ref{sec:2}. In the case of an isolated skyrmion, the radii obtained from $\theta _{\rm LO}(\rho)$ fit the simulation results well, and those calculated obtained from  $\theta _{\rm NNLO}(\rho)$ fit the simulation results  even better. For the skyrmions in the skyrmion phase, their radii are smaller than those of isolated   skyrmions. This is because the skyrmions are now crowded, and are constrained  by the domain walls of the neighbouring skyrmions, consequently the skyrmions shrink.

\begin{figure}
\centering
\includegraphics[width=0.95\textwidth]{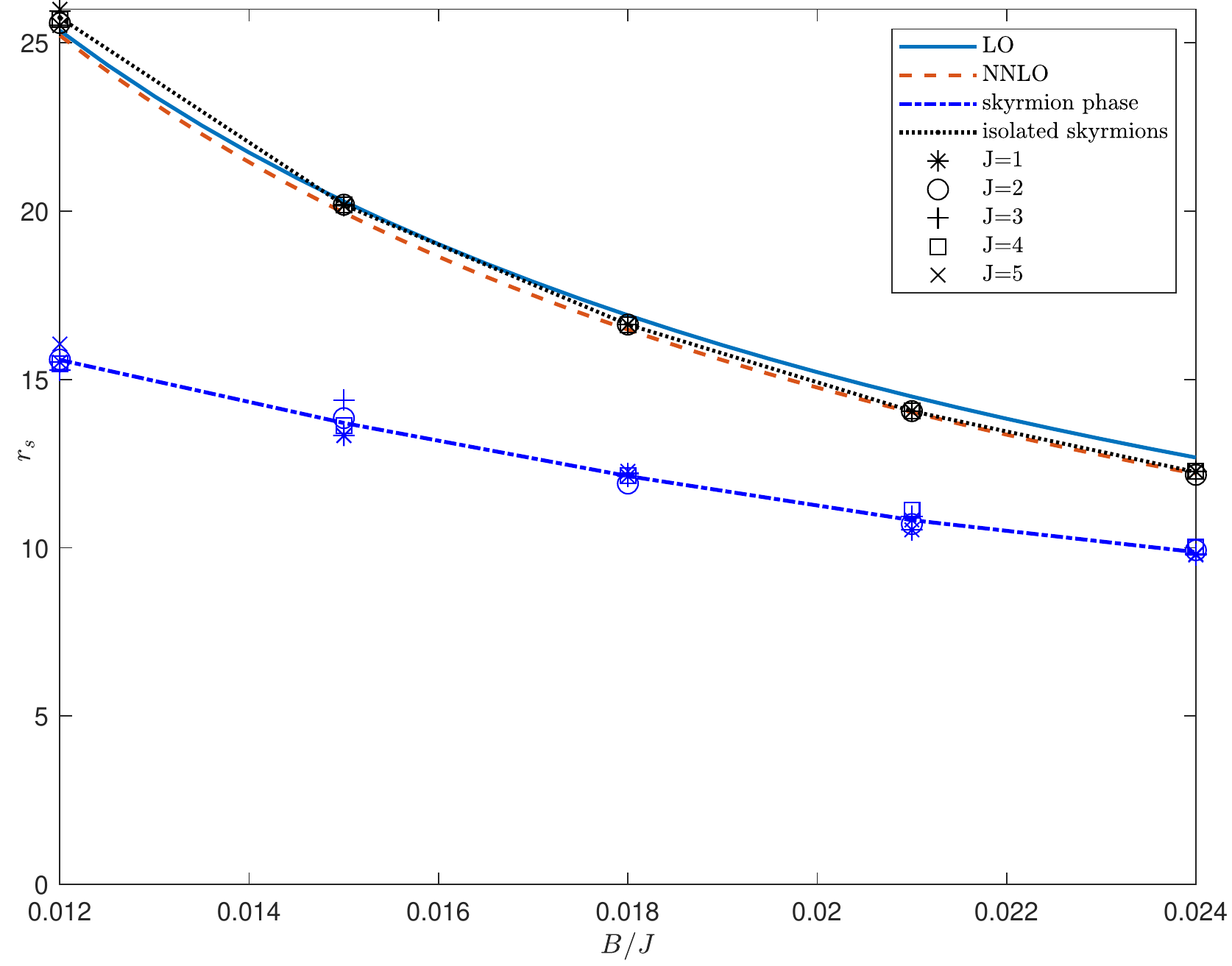}
\caption{The radii $r_s$'s of the skyrmions,   obtained from the simulation and from $\theta _{\rm LO}(\rho)$ and $\theta _{\rm NNLO}(\rho)$. The solid line represents the solution of $\theta _{\rm LO}(\rho)=\cos^{-1}(0.5)$, and the dashed line represents the  solution of $\theta _{\rm NNLO}(\rho)=\cos^{-1}(0.5)$. The dotted line and the dashed dotted line are guides to the eyes.   For an isolated skyrmion, $\theta _{\rm LO}(\rho)$ fits the simulation result well and $\theta _{\rm NNLO}(\rho)$ fits it even  better. For the skyrmions in the skyrmion phase, the radii of the skyrmions are smaller than those of the  isolated skyrmions, because the skyrmions are then crowded and constrained by the neighbouring  skyrmions.}
\label{fig:rs}
\end{figure}

Using dimensionless parameters, $J$ is an energy unit~\cite{JDBRef1,alpha3,bilayer2}, and is also related to the rescaling of the lattice. To make correspondence with  the real material, we use the rescaling method in Refs.~\cite{Tchoe,pin}.
The lattice rescaling factor $r$ is related to helical wavelength $\lambda$ and the lattice spacing $a$ as $r=(D/J)\lambda/\left(2\pi \sqrt{2}a\right)$, and the time unit is rescaled  as $t'=r^2t$.
For example, if  for a real material, we have $\lambda \approx 60\;{\rm nm}$,  $a\approx 4\;{\AA}$ and  $D/J=0.18$, we find $r\approx 3.04$, therefore $\rho=1$ corresponds to $\rho\approx 1.22 {\rm nm}$. Hence  $r_s=15.5176$  corresponds to $ 18.87 {\rm nm}$. The unit of time rescales with the dimensionless $J$ and exchange strength of real material $J'$ as $t'=r^2J\hbar/J'$. For the case of $J=1$, if we adopt $J'=3\;{\rm meV}$, then $t'=r^2J\hbar/J'  \approx 220r^2\;{\rm fs} \approx 0.02\;{\rm ns}$, thus the time step in the simulation corresponds to  $\Delta t\approx 20.3 {\rm fs}$.

\section{\label{sec:4}Applications}

One can predict the skyrmion's behaviour associated with $\theta(\rho)$ by using $\theta_{\rm  LO}(\rho)$ or $\theta_{\rm  NNLO}(\rho)$. Besides, when the relation between the behaviour of the skyrmions and the parameters $J$, $B$ and $D$ can be expressed explicitly, one is able to use such expressions as a guidance  to choose the parameter values of $J$, $B$ and $D$ in experiments. In the following, we give two examples showing that the problems can be greatly simplified by using our method.

\subsection{\label{sec:41} Thiele Equation}

The motion of a skyrmion can be described in terms of  the Thiele equation
\begin{equation}
0=-f_{\rm pin}^{\mu}+g\epsilon ^{\mu\nu}\left(v_{\nu}-j_{\nu}\right)+d^{\mu\nu} \left(\alpha v_{\nu}-j_{\nu}\right),
\label{eq.3.1}
\end{equation}
where $f_{\rm pin}^{\mu}$ is the  pinning force, $g=4\pi Q$  is the gyromagnetic coupling   proportional to the skyrmion number $Q$, $\epsilon^{\mu\nu}$ is the Kronecker tensor, $v$ is the collective velocity of skyrmion and $j$ is the  external electrical current, $\alpha$ is the Gilbert damping coefficient, $d^{\mu\nu}$ is the dissipative matrix given by
\begin{equation}d^{\mu\nu}=\int d^2r \left(\partial _{\mu}\vec{n}\right)\cdot \left(\partial _{\nu}\vec{n}\right).
\end{equation}

From Eq.~(\ref{eq.1.1}), we find
\begin{equation}
\begin{split}
&d^{xy}=d^{yx}=0,\;\;d^{xx}=d^{yy}=2 \pi d_0,\end{split}
\label{eq.3.2}
\end{equation}
with
\begin{equation}
\begin{split}
&d_0\equiv \int _0^{\infty}d\rho\frac{\sin^2\left(\theta(\rho)\right)+\rho^2\left(\frac{\partial \theta(\rho)}{\partial \rho}\right)^2}{2\rho},
\end{split}
\end{equation}
which, according to $\theta_{\rm NNLO}(\rho)$
given in Eq.~(\ref{eq.2.14}),    can be written as
\begin{equation}
\begin{split}
&d_0\approx c_1+c_2b+\mathcal{O}(b^2),\\
\end{split}
\label{eq.3.3}
\end{equation}
where $b$ are parameters in Eq.~(\ref{eq.2.21}), and
\begin{equation}
\begin{split}
&c_1=\int _0^{\infty}dy\left(\frac{\pi ^2 e^{-y} y^2+\sin ^2\left(\pi  e^{-\frac{y}{2}}\right)}{4 y}\right)\approx 2.8991,\\
&c_2=\frac{1}{2}\left(-\text{Ci}(2 \pi )-2 \pi ^2+\gamma _E +\log (2 \pi )\right).\\
\end{split}
\label{eq.3.4}
\end{equation}
Note that $c_1$ is the LO contribution while the $c_2b$ is the NNLO contribution. Using the results of $c_1$, $c_2$ and $b$, we find
\begin{equation}
\begin{split}
&d_0\approx 3.75553+\frac{\frac{B}{J} \left(0.582005 \left(\frac{D}{J}\right)^2-1.75302 \frac{B}{J}\right)}{\left(\frac{B}{J}\right)^2+0.013875 \frac{B}{J} \left(\frac{D}{J}\right)^2-0.114832 \left(\frac{D}{J}\right)^4}.
\end{split}
\label{eq.3.5}
\end{equation}

\begin{figure}
\includegraphics[width=0.9\textwidth]{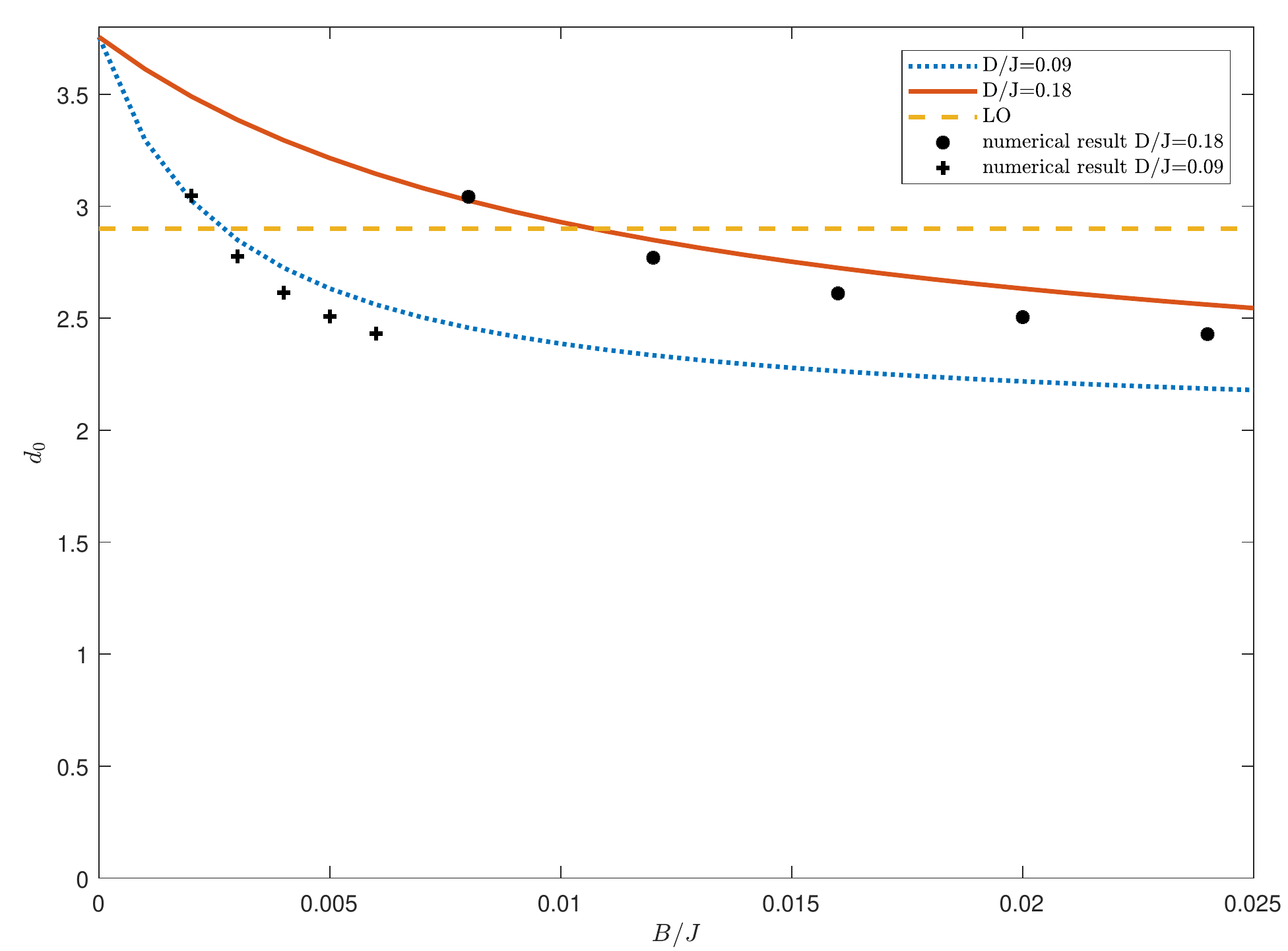}
\caption{\label{fig:5} Comparison of the results on the dissipative matrix element  $d_0$ from our analytical expression  $\theta_{\rm NNLO}$ and from numerical calculation. The dotted line represents the analytical result from $\theta_{\rm NNLO}$ for $D/J=0.09$, the solid line represents  the analytical result from $\theta_{\rm NNLO}$ for  $D/J=0.18$. The plus signs represent numerical results  for $D/J=0.09$ and $B/J=0.002, 0.003, 0.004, 0.005, 0.006$. The filled circles represent  numerical results  for  $D/J=0.18$ and $B/J=0.008, 0.012, 0.016, 0.020, 0.024$. The dashed line represents  the analytical result from $\theta_{\rm LO}$.}
\end{figure}

\begin{figure}
\includegraphics[width=0.9\textwidth]{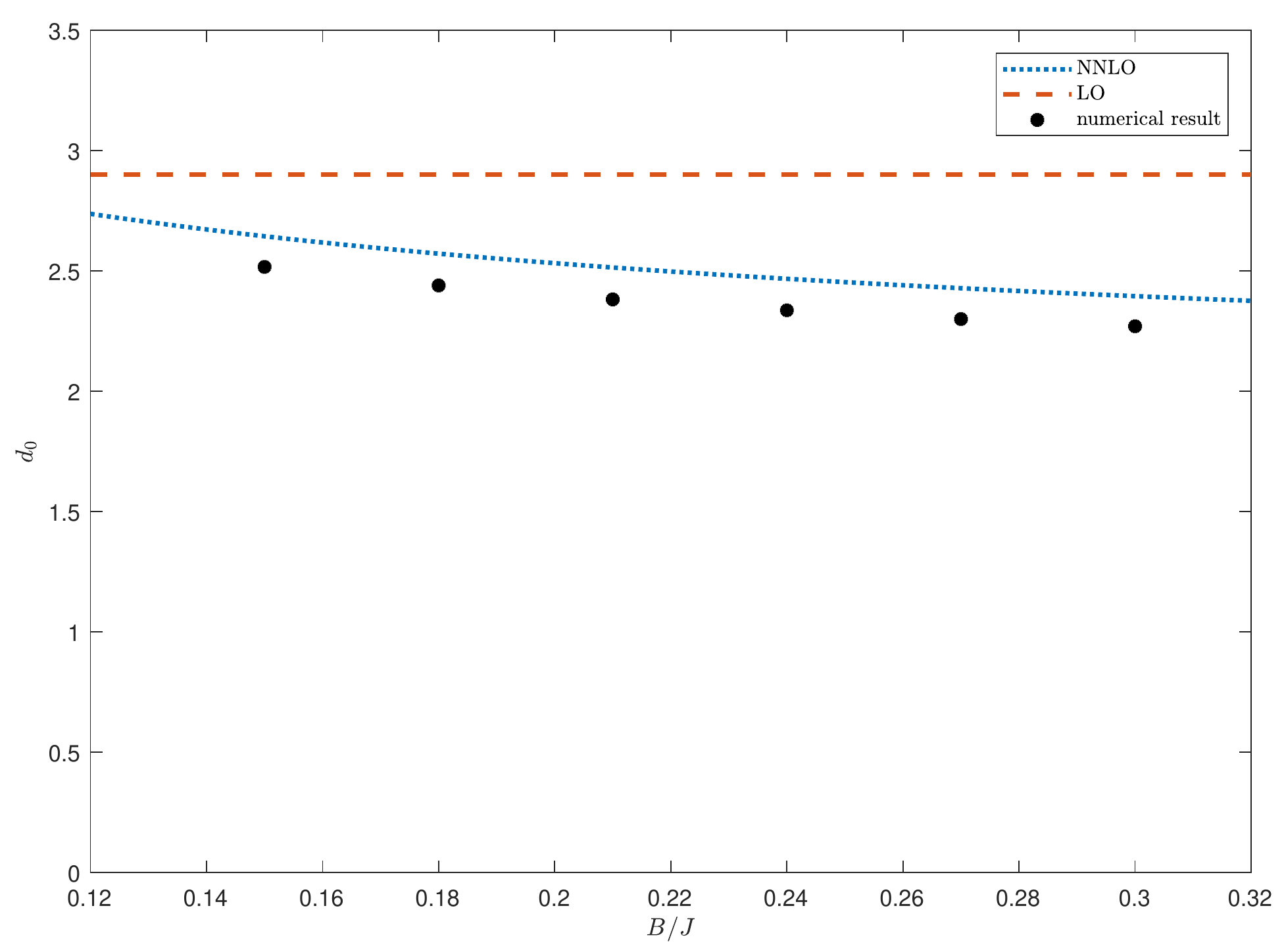}
\caption{\label{fig:6} Comparison of the results on the dissipative matrix element  $d_0$ from our analytical expression  $\theta_{\rm NNLO}$ and from numerical calculation. The dotted line represents the analytical result from $\theta_{\rm NNLO}$ for $D/J=0.5$.  The filled circles represent  numerical results for $D/J=0.5$ and $B/J=0.15, 0.18, 0.21, 0.24, 0.27, 0.3$.   The dashed line represents  the analytical result  from $\theta_{\rm LO}$. }
\end{figure}

Now we compare this result with the numerical result, as shown in Figs.~\ref{fig:5} and \ref{fig:6}, for the parameter regimes studied in Refs.~\cite{ThieleStudyAndJDB,JDBRef1,JDBRef2}. We find that our variational calculation  agrees with  the numerical result  well. The discrepancy  increases with the deviation of $d_0$ from $c_1$, because as $\theta(\rho)$ deviates from $\phi _0$, the higher order terms become important.

\subsection{\label{sec:42}Interaction of skyrmions in a bilayer system}

The interaction between two skyrmions on two separated and overlapped planes is an interesting problem studied in Ref.~\cite{bilayer2} using micromagnetic simulations as well as the analysis based on Thiele equation. When two skyrmions are close to each other, $\theta(\rho)$ plays an important role in the interaction between the two skyrmions.

The normalized potential between two skyrmions can be written as~\cite{bilayer2}
\begin{equation}
\begin{split}
&u({\bf r}_d)=\int d^2 r\left(1-{\rm sign }(J_{\rm inter}) {\bf n}_1({\bf r}-\frac{{\bf r}_d}{2})\cdot {\bf n}_2({\bf r}+\frac{{\bf r}_d}{2})\right),\\
\end{split}
\label{eq.3.6}
\end{equation}
where ${\bf n}_1$ and ${\bf n}_2$ are magnetic moments of the two skyrmions, $J_{\rm inter}$ is the Heisenberg coupling  between two layers, ${\bf r}_d$ is the distance between the two skyrmions. Suppose the DM interaction strengths of the two skrymions are  $D_1$ and $D_2$, respectively.

To study the case of  $D_1=-D_2$, one needs to consider a more general  parameterization~\cite{nagaosa}
\begin{equation}
\begin{split}
&{\bf n}({\bf r})=\left(\cos(\gamma +m \phi)\sin(\theta(\rho)),\sin(\gamma +m \phi)\sin(\theta(\rho)),g \cos (\theta (\rho))\right),\\
\end{split}
\end{equation}
where $(\phi, \rho)$ is ${\bf r}$ in polar coordinates, $\gamma$ is the helicity angle, $m= 1$ for a skyrmion  and  $m= -1$ for an anti-skyrmion, $g=\pm 1$. The skyrmion number is
\begin{equation}
\begin{split}
&Q=\frac{1}{4\pi}\int dxdy {\bf n}\cdot \left(\frac{\partial {\bf n}}{\partial x}\times \frac{\partial {\bf n}}{\partial y}\right)=-mg.\\
\end{split}
\end{equation}

With isotropic DM interaction, we consider the case of  $m=1$~\cite{antiskyrmion}. The Euler-Lagrange equation yields
\begin{equation}
\begin{split}
&\frac{ \sin(\theta)\cos(\theta)}{\rho}-\theta ' - \rho \theta ''-2\frac{\hat{D}}{J}\sin^2(\theta)+\frac{\hat{B}}{J} \rho \sin (\theta)=0,
\end{split}
\end{equation}
with $\hat{D} \equiv g\sin(\gamma)D$, $\hat{B}\equiv gB$. The parameters in Eq.~(\ref{eq.1.1}) correspond to  $g=1$ and $\gamma = \pi / 2$ here.  $D<0$ case in Ref.~\cite{bilayer2} correspond to $g=1$ and $\gamma = 3\pi / 2$. For $D_2=-D_1$, by setting $\gamma =3\pi /2$, the equation of motion of $\theta(\rho)$ is the same as that for $D_1$ with $\gamma =\pi /2$.
As a result, in considering the two interacting skrymions, with  $D_1=D_2$ or $D_1=-D_2$, we have two skyrmions with the same size and same shape $\theta(\rho)$. Suppose the intralayer Hamiltonians dominate the interlayer interaction,  the effect of inter-skyrmion interaction on $\theta(\rho)$ can be neglected. Consequently, one can use the functions  $\theta_{\rm LO}$ and $\theta_{\rm NNLO}$ above for each skyrmion.

For $D_1=\pm D_2$, we define the corresponding normalized potential as $u_{\pm}({\bf r}_d)$,
\begin{equation}
\begin{split}
&u_{\pm}({\bf r}_d)=\int d^2 r\left[1-\cos \left(\theta(|{\bf r}+\frac{{\bf r}_d}{2}|)\right) \cos \left(\theta(|{\bf r}-\frac{{\bf r}_d}{2}|)\right)\right.\\
&\left.\mp\frac{\left(r^2-\frac{{\bf r}_d^2}{4}\right) \sin \left(\theta(|{\bf r}+\frac{{\bf r}_d}{2}|)\right)\sin \left(\theta(|{\bf r}-\frac{{\bf r}_d}{2}|)\right)}{|{\bf r}+\frac{{\bf r}_d}{2}||{\bf r}-\frac{{\bf r}_d}{2}|}\right].\\
\end{split}
\label{eq.3.8}
\end{equation}

Note that both $\theta_{\rm LO} (\rho)$ and $\theta_{\rm NNLO} (\rho)$ are functions of $-\omega \rho^2$ and can be written as $\theta '(-\omega\rho^2/2)$, for simplicity, we can take the ${\bf x}$-axis as the direction of ${\bf r}_d$ and rewrite $u_{\pm}$ as
\begin{equation}
\begin{split}
&u_{\pm}(r_d)=\frac{1}{\omega}\bar{u}_{\pm}
(\frac{\sqrt{\omega}r_d}{2}),\\
\end{split}
\label{eq.3.10}
\end{equation}
where
\begin{equation}
\begin{split}
&\bar{u}_{\pm}(r'_d)=\int dxdy\left[1-\cos \left(\theta'(-\frac{1}{2}((x+r'_d)^2+y^2))\right) \cos \left(-\frac{1}{2}((x-r'_d)^2+y^2))\right)\right.\\
&\left.\mp\frac{\left(x^2+y^2-(r'_d)^2\right) \sin \left(\theta'(-\frac{1}{2}((x+r'_d)^2+y^2))\right)\sin \left(\theta'(-\frac{1}{2}((x-r'_d)^2+y^2))\right)}{\sqrt{(x+r'_d)^2+y^2}\sqrt{(x-r'_d)^2+y^2}}\right],\\
\end{split}
\label{eq.3.11}
\end{equation}
with $2r'_d$ being the rescaled dimensionless distance of the two skyrmions.  $\bar{u}_{\pm}(r'_d)$ is   a function independent of $J$, $B$ and $D$, and can be used  for various values of $J$, $B$ and $D$.

It is difficult to obtain the analytical result of $\bar{u}_{\pm}(r_d)$, so we use Pad${\rm \acute{e}}$ approximants~\cite{pade,padeapp}. When the two skyrmions are far away from each other, the skyrmions should be independent of each other, so  the functions $\bar{u}_{\pm}(r_d')$ should  asymptotically become constants. Numerically we find $|\partial _{r_d}\bar{u}_{\pm}(3)|<0.02$, so we use k-points $[m,n]$ order Pad${\rm \acute{e}}$ approximation to write  $\bar{u}_{\pm}(r_d')$   as
\begin{equation}
\begin{split}
&\bar{u}_{\pm}(r'_d)\approx \left\{\begin{array}{c}\tilde{u}_{\pm}(r'_d),\;\;0\le r'_d \le 3; \\ \tilde{u}_{\pm}(3),\;\; r'_d > 3,
\end{array}
\right.
\end{split}
\label{eq.3.12}
\end{equation}
where
\begin{equation}
\begin{split}
&\tilde{u}_{\pm}(r'_d)=\frac{\sum _{i=0}^mp_i{r'_d}^i}{1+\sum _{j=1}^nq_j{r'_d}^j}.
\end{split}
\end{equation}

We use three-point $[5,4]$ order Pad${\rm \acute{e}}$ approximants where $p_{0,\ldots,4}$ and $q_{1,\ldots,4}$ are $9$ constants independent of $J$, $D$ and $B$ and can be determined from equations
\begin{equation}
\begin{split}
&\tilde{u}_{\pm}(r'_d)-\bar{u}_{\pm}(r'_d)=0+\mathcal{O}\left((r'_d-r'_{k})^l\right),\\
\end{split}
\label{eq.3.13}
\end{equation}
where $k=1,2,3$,$r'_{1,2,3}=0,3/2,3$, $l=3$. Using $\theta _{\rm LO}(\rho)$, we find
\begin{equation}
\begin{split}
&u_{\pm}(r_d)\approx \left\{\begin{array}{c}u_{\rm \pm LO}(r_d),\;\;0\le r_d\le r_{d\;max};\\u_{\rm\pm LO}(r_{d\;max}),\;\;r_d>r_{d\;max}.\end{array} \right.\\
\end{split}
\label{eq.3.14}
\end{equation}
with $r_{d\;max}=6/\sqrt{\omega _{\rm LO}}$,
\begin{equation}
\begin{split}
&u_{\rm +LO}(r_d)= \frac{D^2 r_d^2 \left(0.625637 B^2 r_d^2-4.20541 B D r_d+9.10783 D^2\right)}{0.0240738 B^4 r_d^4-0.177479 B^3 D r_d^3+0.532727 B^2 D^2 r_d^2-0.751001 B D^3 r_d+D^4},\\
&u_{\rm -LO}(r_d)= \frac{D^2 \left(25.9614 B^4 r_d^4-281.728 B^3 D r_d^3+1249.66 B^2 D^2 r_d^2-3067.97 B D^3 r_d+5066.63 D^4\right)}{B^2 \left(B^4 r_d^4-11.4014 B^3 D r_d^3+55.5842 B^2 D^2 r_d^2-153.947 B D^3 r_d+254.237 D^4\right)}.\\
\end{split}
\label{eq.3.15}
\end{equation}

\begin{figure}
\begin{overpic}[width=0.8\textwidth]{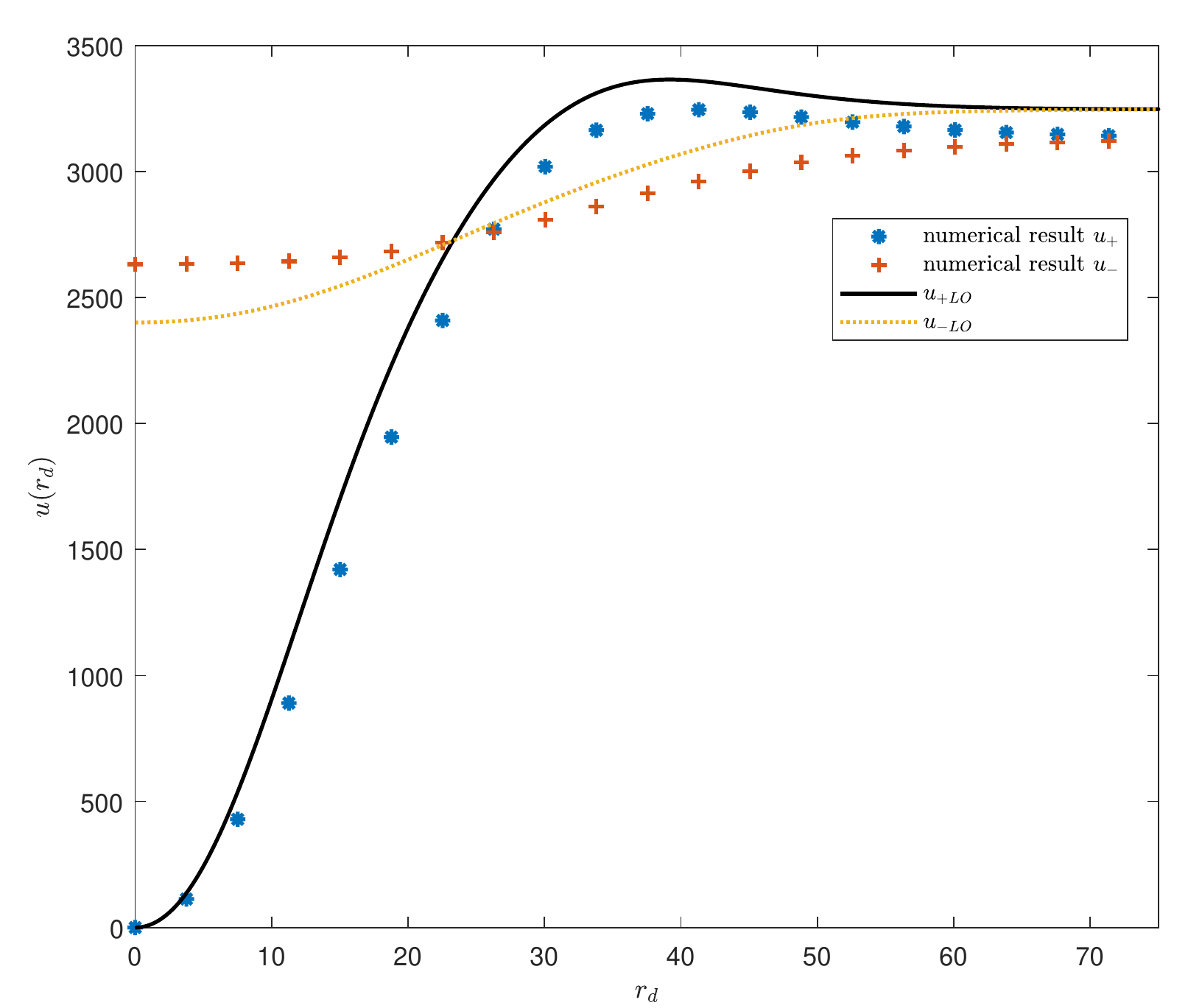}
 \put(40,10){\includegraphics[width=0.43\textwidth]{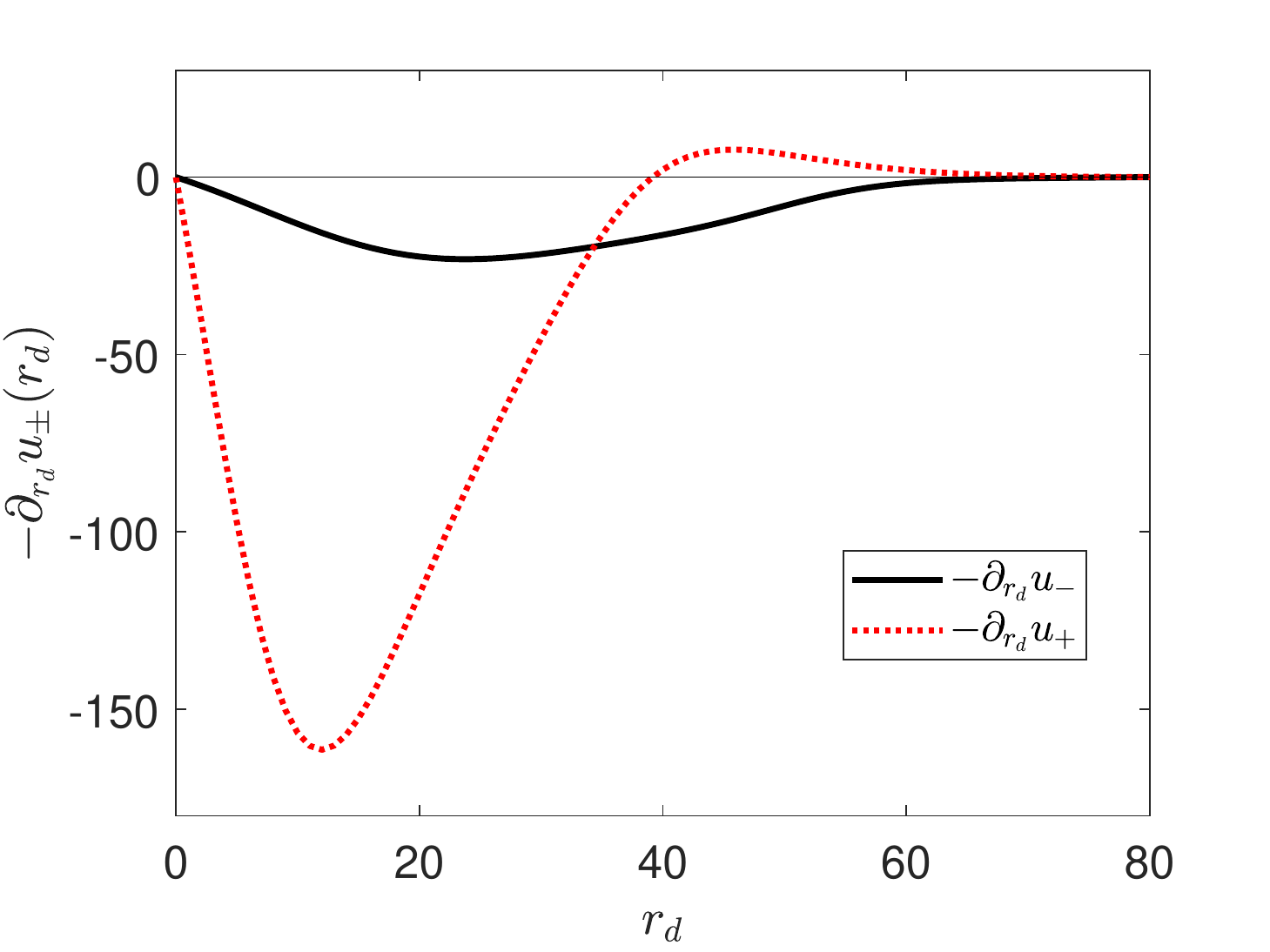}}
\end{overpic}
\caption{\label{fig:7} The normalized potentials obtained from $\theta_{\rm LO}$ and from numerical calculation. The solid line represents $u_{\rm + LO}(r_d)$, while the stars represent $u_+(r_d)$ from numerical calculation.   The dashed line represents $u_{\rm - LO}(r_d)$, while the plus dots represent $u_-(r_d)$ from numerical calculation.   The results are given  for $D/J=0.18$ and $B/J=0.0164$. When $r_d>r_{d\;max}\approx 75.12$, $u_{\pm}(r_d)\approx   u_{\rm LO}(r_{d\;max})$ is about a constant. The interaction $-\partial _{r_d}u$ between two skyrmions are also determined. $-\partial _{r_d}u <0$ means attractive force between the skyrmions, while $-\partial _{r_d}u >0$ means repulsive force  between the skyrmions.}
\end{figure}

We also calculate $u_{\pm}(r_d)$ using the numerical result of $\theta(\rho)$  for $D/J=0.18$ and $B/J=0.0164$, in the regime of the skyrmion phase studied in Ref.~\cite{ThieleStudyAndJDB}. The numerical result and the leading order approximation are shown in Fig.~\ref{fig:7}. Up to LO, our analytical result fits the numerical result well. It is also very convenient to obtain the interaction between the two skyrmions using $F(r_d)= -|J_{\rm inter}|\partial _{r_d} u(r_d)$~\cite{bilayer2} and Eq.~(\ref{eq.3.15}). The function  $-\partial _{r_d}u_{\rm \pm LO}(r_d)$ is also shown in Fig.~\ref{fig:7}.

\section{\label{sec:5}Summary}

As a topological soliton, a skyrmion is often treated as a point-like particle on large scales. However, when the scale of the dynamics is comparable with the radius of the  skyrmion, we need to consider the shape of the skyrmion. Moreover, we need $\theta(\rho)$ to determine  the dissipative matrix in the Thiele equation. Hence the study of the $\theta(\rho)$ is both interesting and important.

In this paper, we propose a method to represent $\theta (\rho)$ approximately yet efficiently in terms of the eigenfunctions of Schr\"{o}dinger operator of the harmonic oscillator. Using variational approach, we find that the result can be written as a superposition of the first few eigenfunctions Eqs.~(\ref{eq.2.14}), and we have determined that superposition coefficients as well as the frequency of the harmonic oscillator as a function of the parameters of the magnetic system. Using the result, we are  immediately able to calculate the radii of the skrymions, which are verified by the numerical calculation and also by lattice  simulation. We also use our method to study the dissipative matrix in the Thiele equation. We obtain the matrix elements as explicit functions of the  parameters $B$, $J$ and $D$, which agree with the numerical results. We also use our method to study the interaction between two skyrmions on two layers, with result confirmed by the numerical calculation.

This work is supported by National Natural Science Foundation of China (Grant No. 11374060 and No. 11574054).

\clearpage

\appendix

\beginsupplement

\section{\label{sec:a}Supplemental information}

\subsection{\label{sec:expand}Ansatz functions as examples of harmonic oscillator expansion}

One can choose any complete set of orthogonal functions to expand $\theta (\rho)$. For better convergence, we choose the eigenfunctions of harmonic oscillator such that the LO wave function is close to the numerical results. As examples, we obtain the harmonic oscillator expansions of  the ansatz functions of $\theta (\rho)$ used in previous works, in which $\theta(\rho)$ is linear in $\rho$~\cite{Jiang,NParameterAndLinear,thetarho}, or satisfy $\theta(\rho) =\arctan \left(\exp\left(-\rho/\Delta\right)\right)$~\cite{NParameterArcTan,arctan}.

\begin{figure}
\subfloat[$f_1(x)$]{\includegraphics[width=0.5\textwidth]{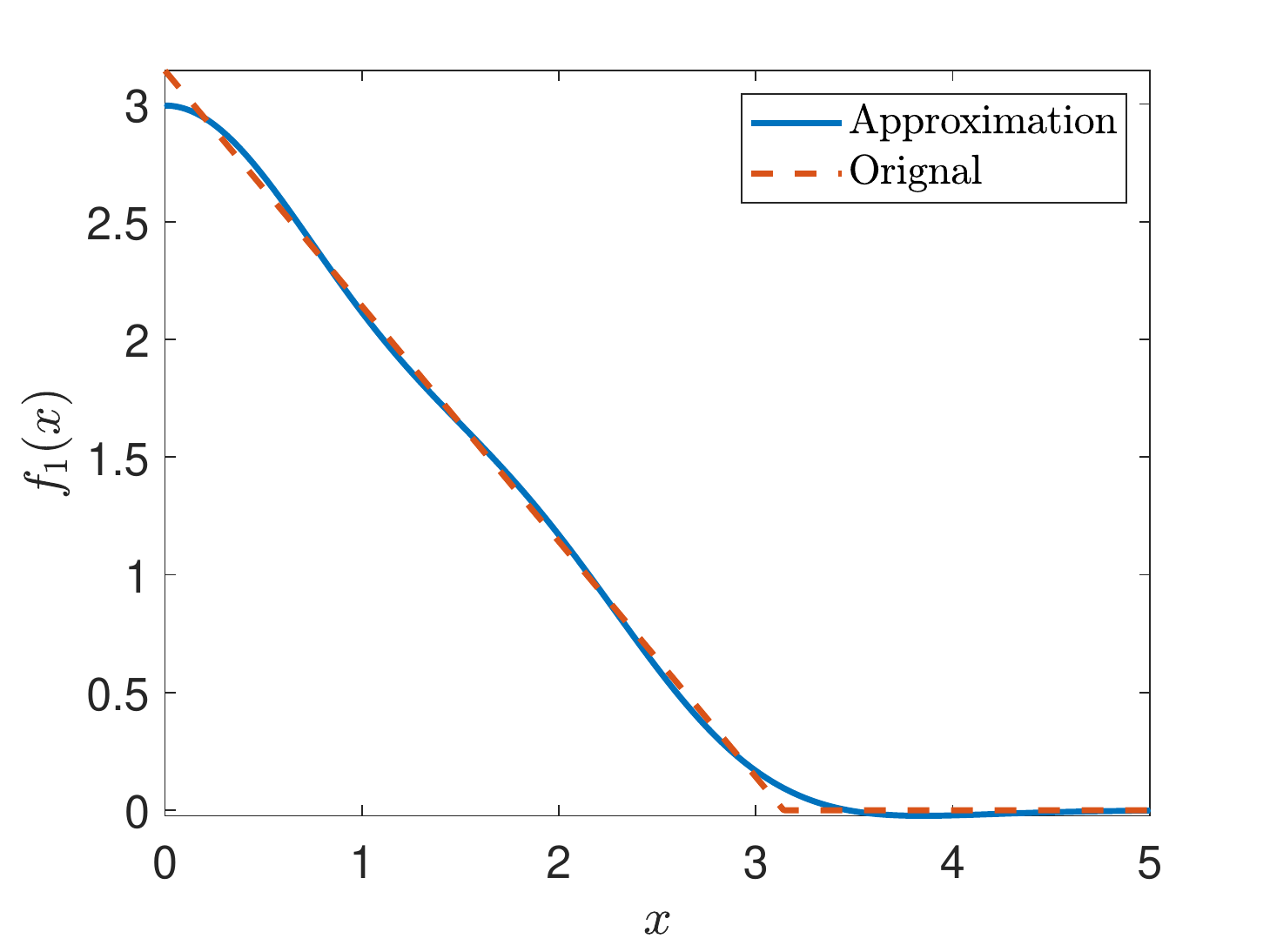}}\hfill
\subfloat[$f_2(x)$]{\includegraphics[width=0.5\textwidth]{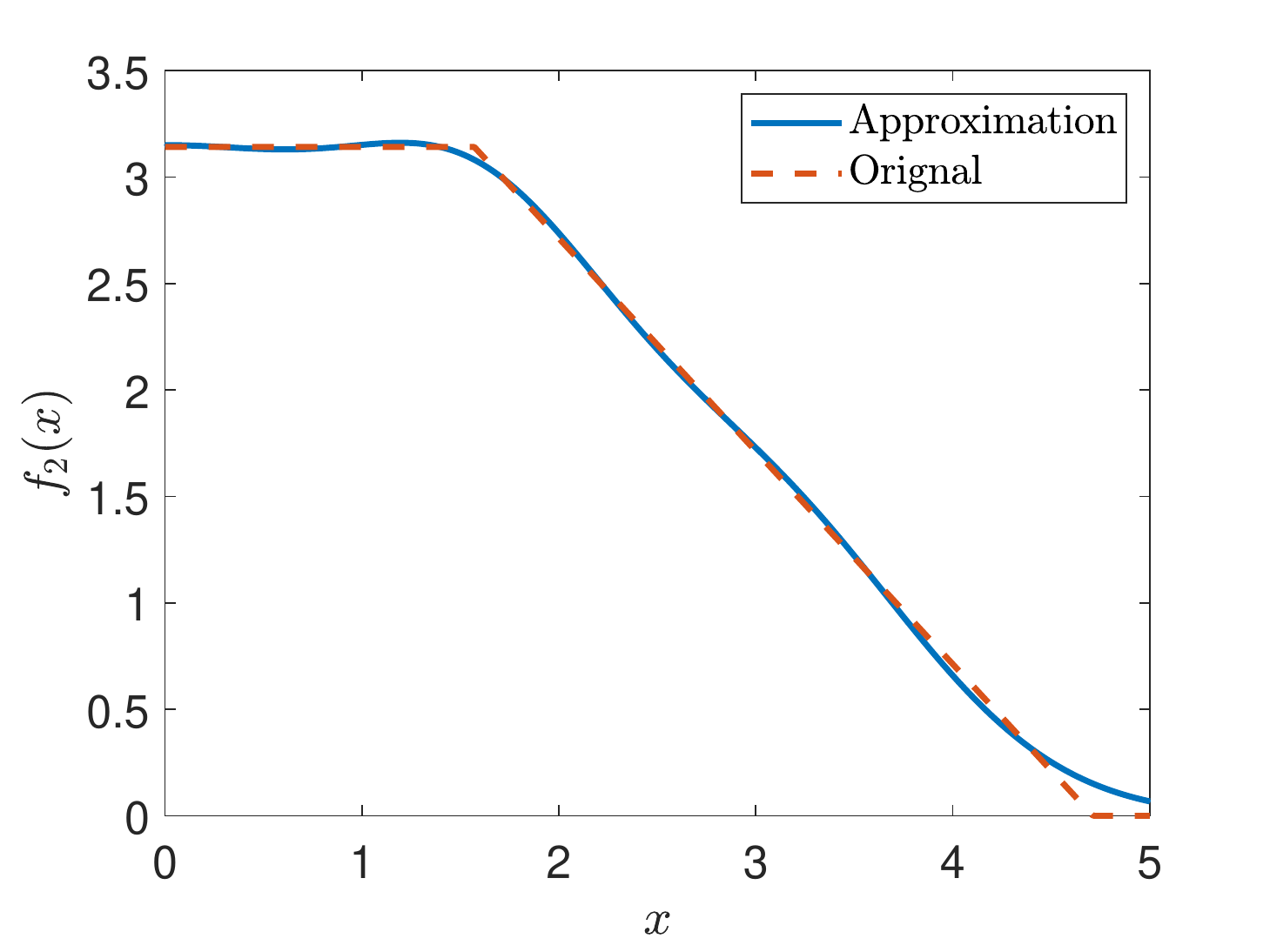}}\vfill
\subfloat[$f_3(x)$]{\includegraphics[width=0.5\textwidth]{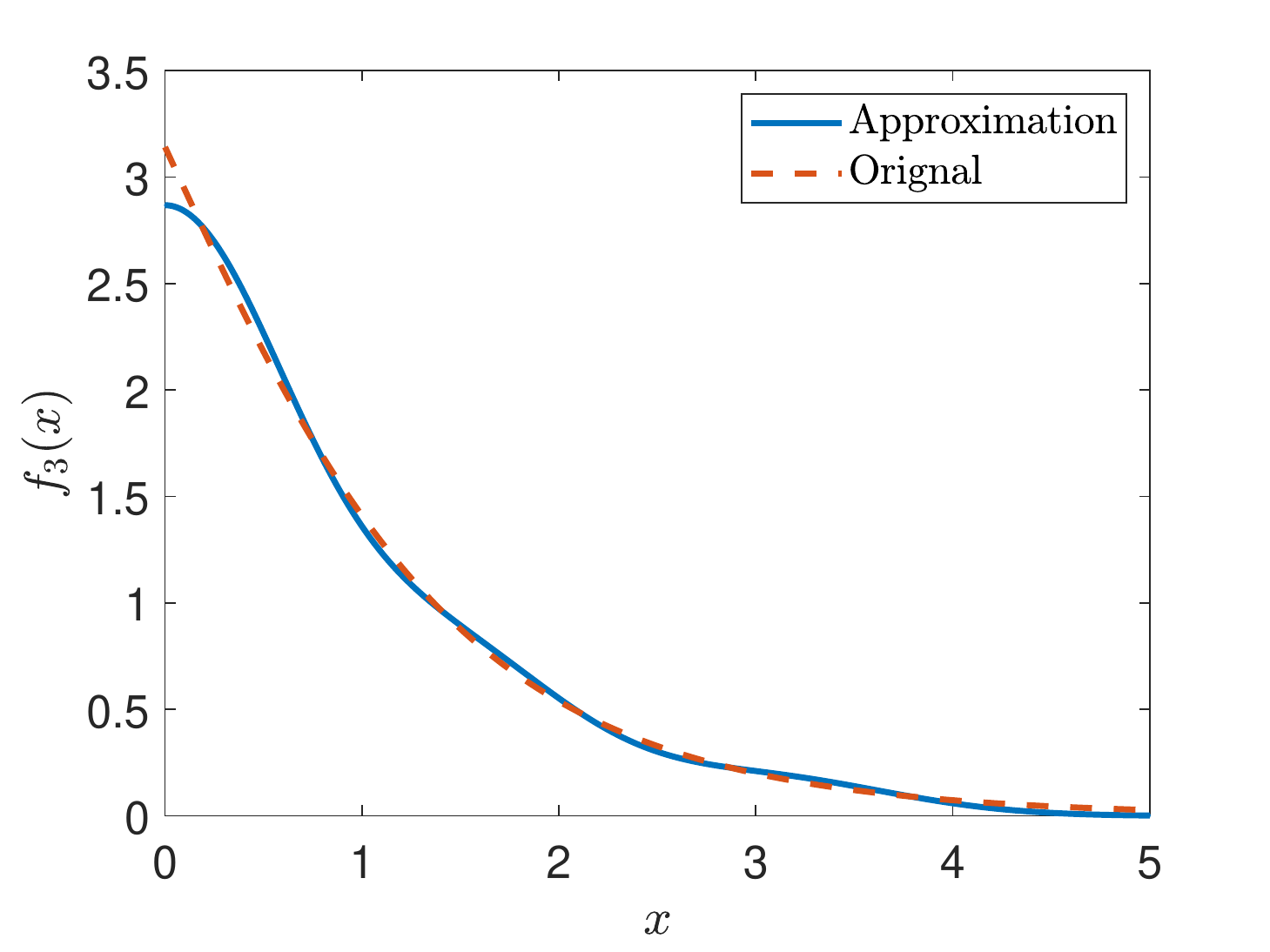}}\hfill
\subfloat[$f_4(x)$]{\includegraphics[width=0.5\textwidth]{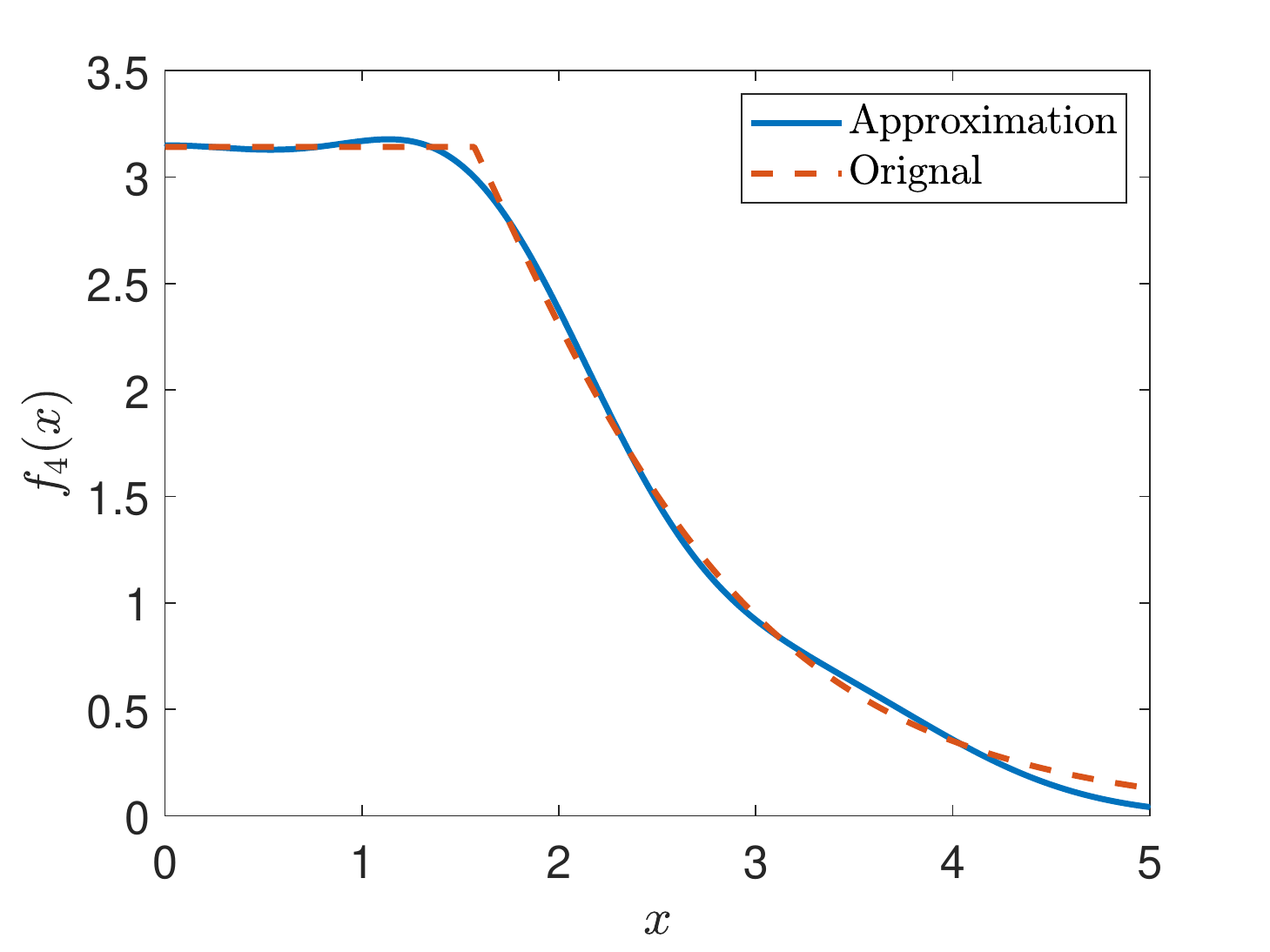}}\vfill
\caption{ The harmonic oscillator expansion  of some functions  listed in Table.~\ref{tab:1}.  $\omega$ is arbitrarily set to be $1$. In each plot, the dashed line is for the function while the solid line is for the  expansion $\sum _{n=0}^5C_n\phi _{2n,\omega}(x)$. These functions have been used for  $\theta (\rho)$ previously. }
\label{fig:2}
\end{figure}

The results  are listed  in Table.~\ref{tab:1} and  shown in  Fig.~\ref{fig:2}, with $\omega$ arbitrarily set to be $1$. Among these examples, the functions $f_1(x)$ and $f_2(x)$ are the ansatz functions for $\theta (\rho)$ in Ref.~\cite{Jiang,NParameterAndLinear,NParameterArcTan,thetarho}, while the functions $f_3(x)$ and $f_4(x)$ are the ansatz functions for $\theta (\rho)$ in Ref.~\cite{arctan,NParameterArcTan}. It can be seen that the expansions approximate the original functions very well.

\begin{table}
\caption{ The harmonic oscillator expansion of some functions, which were previously used for $\theta(\rho)$, and are also shown in Fig.~\ref{fig:2}.   $\omega$ is arbitrarily set to be 1.    }
\label{tab:1}
\begin{tabular}{|c|c|c|c|c|c|c|}
\hline
\hline
$f(x)\approx \sum _{\substack{n=0}}^5C_n\phi _{2n,\omega}(x)$&$C_0$&$C_1$&$C_2$&$C_3$&$C_4$&$C_5$\\
\hline
\hline
$f_1(x)=\left\{\begin{array}{c}\pi-x,\;\;x\in [0,\pi];\\0,\;\; x>\pi.\end{array}\right.$&4.414&1.012&0.354&-0.173&-0.042&0.009\\
\hline
$f_2(x)=\left\{\begin{array}{c}\pi,\;\;x\in [0,\frac{\pi}{2});\\ \frac{3\pi}{2}-x,\;x \in [\frac{\pi}{2},\frac{3\pi}{2}],\\ 0,\;x>\frac{3\pi}{2}.\end{array}\right.$&5.821&3.498&1.969&1.101&0.650&0.168\\
\hline
$f_3(x)=4\tan^{-1}(\exp(-x))$&3.486&0.069&0.428&-0.031&0.157&-0.040\\
\hline
$f_4(x)=\left\{\begin{array}{c}\pi,\;\;x\in [0,\frac{\pi}{2}];\\4\tan^{-1}\left(e^{-(x-\frac{\pi}{2})}\right),\;\; x>\frac{\pi}{2}.\end{array}\right.$&5.746&3.000&1.014&0.371&0.407&0.117\\
\hline
\hline
\end{tabular}
\end{table}

In the above examples, an arbitrary value of $\omega$ was used. However, an appropriate value of $\omega$ leads to better convergence.

\subsection{\label{sec:solution}The solution at NNLO}

Substitute the trial solution Eq.~(\ref{eq.2.14}) for  $F$ in Eqs.~(\ref{eq.2.2}) and (\ref{eq.2.3}). By minimizing   $F$, one obtains three equations  with three undetermined variables.

The equation $dF/db=0$ leads to Eq.~(\ref{eq.2.15}) with
\begin{equation}
\begin{split}
&a_{111}=-\text{Ci}(2 \pi )-2 \pi ^2+\gamma _E +\log (2 \pi )\approx -17.3016,\\
&a_{112}=2 \pi ^2 \left(\; _3F_4\left(\left.\begin{array}{c}1,1,1\\ \frac{1}{2},2,2,2\end{array}\right|-\pi ^2\right)+2\right)\approx 39.1263,\\
&a_{113}=6 \pi ^2 \left(\; _4F_5\left(\left.\begin{array}{c}1,1,1,1\\ \frac{1}{2},2,2,2,2\end{array}\right|-\pi ^2\right)+4\right) \approx 255.4126,\\
&a_{211}=0,\;\;\;a_{212}=-a_{111},\;\;\;a_{213}=a_{112},\;\;a_{311}=a_{312}=a_{313}=0,\\
\end{split}
\label{eq.2.16}
\end{equation}
while $dF/da=0$ leads to
\begin{equation}
\begin{split}
&a_{121}=\int _0^{\infty}dy\left(\pi  e^{-\frac{y}{2}} y^{3/2} \left(-y+\cos \left(2 \pi  e^{-\frac{y}{2}}\right)+4\right)\right)\approx -13.4774,\\
&a_{122}=\int _0^{\infty}dy\left(-2 \left(\pi ^2 e^{-y} y^{5/2} \sin \left(2 \pi  e^{-\frac{y}{2}}\right)\right)\right)\approx -36.5000,\\
&a_{123}=\int _0^{\infty}dy\left(-2 \left(\pi ^2 e^{-y} y^{7/2} \sin \left(2 \pi  e^{-\frac{y}{2}}\right)\right)\right)\approx -129.3094,\\
&a_{221}=\int _0^{\infty}dy\left(\pi  e^{-\frac{y}{2}} \sqrt{y} \left(-y+\cos \left(2 \pi  e^{-\frac{y}{2}}\right)+2\right)\right)\approx -7.5206,\\
&a_{222}=\int _0^{\infty}dy\left(-2 \left(\pi ^2 e^{-y} y^{3/2} \sin \left(2 \pi  e^{-\frac{y}{2}}\right)\right)\right)\approx -9.0845,\\
&a_{223}=a_{122},\;\;\;a_{321}=a_{322}=-\frac{1}{2}a_{221},\;\;\;a_{323}=-\frac{1}{2}a_{121},\\
\end{split}
\label{eq.2.17}
\end{equation}
and $dF/d\omega=0$ leads to
\begin{equation}
\begin{split}
&a_{131}=-2 \pi ^2 \; _3F_4\left(\left.\begin{array}{c}1,1,1\\ \frac{3}{2},2,2,2\end{array}\right|-\frac{\pi ^2}{4}\right)\approx -16.2190,\\
&a_{132}=-6 \pi ^2 \; _4F_5\left(\left.\begin{array}{c}1,1,1,1\\ \frac{1}{2},2,2,2,2\end{array}\right|-\frac{\pi ^2}{4}\right)\approx -43.6330,\\
&a_{133}=-24 \pi ^2 \; _5F_6\left(\left.\begin{array}{c}1,1,1,1,1\\ \frac{1}{2},2,2,2,2,2\end{array}\right|-\frac{\pi ^2}{4}\right)\approx -204.0066,\\
&a_{231}=-4 (-\text{Ci}(\pi )+\gamma _E +\log (\pi )) \approx -6.5931,\\
&a_{232}=-2 \pi ^2 \; _3F_4\left(\left.\begin{array}{c}1,1,1\\ \frac{1}{2},2,2,2\end{array}\right|-\frac{\pi ^2}{4}\right)\approx -10.1534,\\
&a_{233}=a_{132},\;\;a_{331}=-\frac{1}{2}a_{231},\;\;a_{332}=-a_{231},\;\;a_{333}=-a_{131},\\
\end{split}
\label{eq.2.18}
\end{equation}
where $_p F_q$ are hypergeometric functions.

The solution can be written as
\begin{equation}
\begin{split}
&a\approx\frac{k_3 k_4-k_1 k_6}{k_2 k_6-k_3 k_5},\;\;\;\;b\approx\frac{k_2k_4-k_1k_5}{k_3k_5-k_2k_6},\\
&\sqrt{\omega}\approx \frac{B (a_{331}+a_{332}a+a_{333} b)}{D (a_{321}+a_{322} a+a_{323} b)},\\
\end{split}
\label{eq.2.19}
\end{equation}
with
\begin{equation}
\begin{split}
&k_1=\frac{1}{2} \left(\frac{\left(\frac{D}{J}\right)^2 a_{321} (a_{121} a_{331}-a_{131} a_{321})}{\frac{B}{J} a_{331}^2}+a_{111}\right),\\
&k_2=\frac{1}{2} \left\{\frac{\left(\frac{D}{J}\right)^2}{\frac{B}{J} a_{331}^3} \left[a_{331} \left(a_{121}a_{322} a_{331}+a_{122}a_{321}a_{331}-2 a_{131}a_{321}a_{322} -a_{132} a_{321}^2\right)\right.\right.\\
&\left.\left.+a_{321}a_{332} (2 a_{131}a_{321}-a_{121}a_{331})\right]+a_{112}\right\},\\
&k_3=\frac{1}{2} \left\{\frac{\left(\frac{D}{J}\right)^2}{\frac{B}{J} a_{331}^3}\left[a_{331}\left(a_{121}a_{323}a_{331}+a_{123}a_{321}a_{331}-2a_{131}a_{321}a_{323}-a_{133}a_{321}^2\right)\right.\right.\\
&\left.\left.+a_{321}a_{333} (2 a_{131}a_{321}-a_{121}a_{331})\right]+a_{113}\right\},\\
&k_4=\frac{\left(\frac{D}{J}\right)^2 a_{321} (a_{221} a_{331}-a_{231}a_{321})}{2 \frac{B}{J} a_{331}^2},\\
&k_5=\frac{1}{2} \left\{\frac{\left(\frac{D}{J}\right)^2}{\frac{B}{J} a_{331}^3} \left[a_{331} \left(a_{221}a_{322} a_{331}+a_{222}a_{321}a_{331}-2 a_{231}a_{321}a_{322} -a_{232} a_{321}^2\right)\right.\right.\\
&\left.\left.+a_{321}a_{332} (2 a_{231}a_{321}-a_{221}a_{331})\right]+a_{212}\right\},\\
&k_6=\frac{1}{2} \left\{\frac{\left(\frac{D}{J}\right)^2}{\frac{B}{J} a_{331}^3}\left[a_{331}\left(a_{221}a_{323}a_{331}+a_{223}a_{321}a_{331}-2a_{231}a_{321}a_{323}-a_{233}a_{321}^2\right)\right.\right.\\
&\left.\left.+a_{321}a_{333} (2 a_{231}a_{321}-a_{221}a_{331})\right]+a_{213}\right\}.\\
\end{split}
\label{eq.2.20}
\end{equation}

\subsection{\label{sec:simulation}Radii of the skyrmions in simulations}

\begin{figure}
\subfloat[$J=3,B/J=0.012$]{\includegraphics[width=0.49\textwidth]{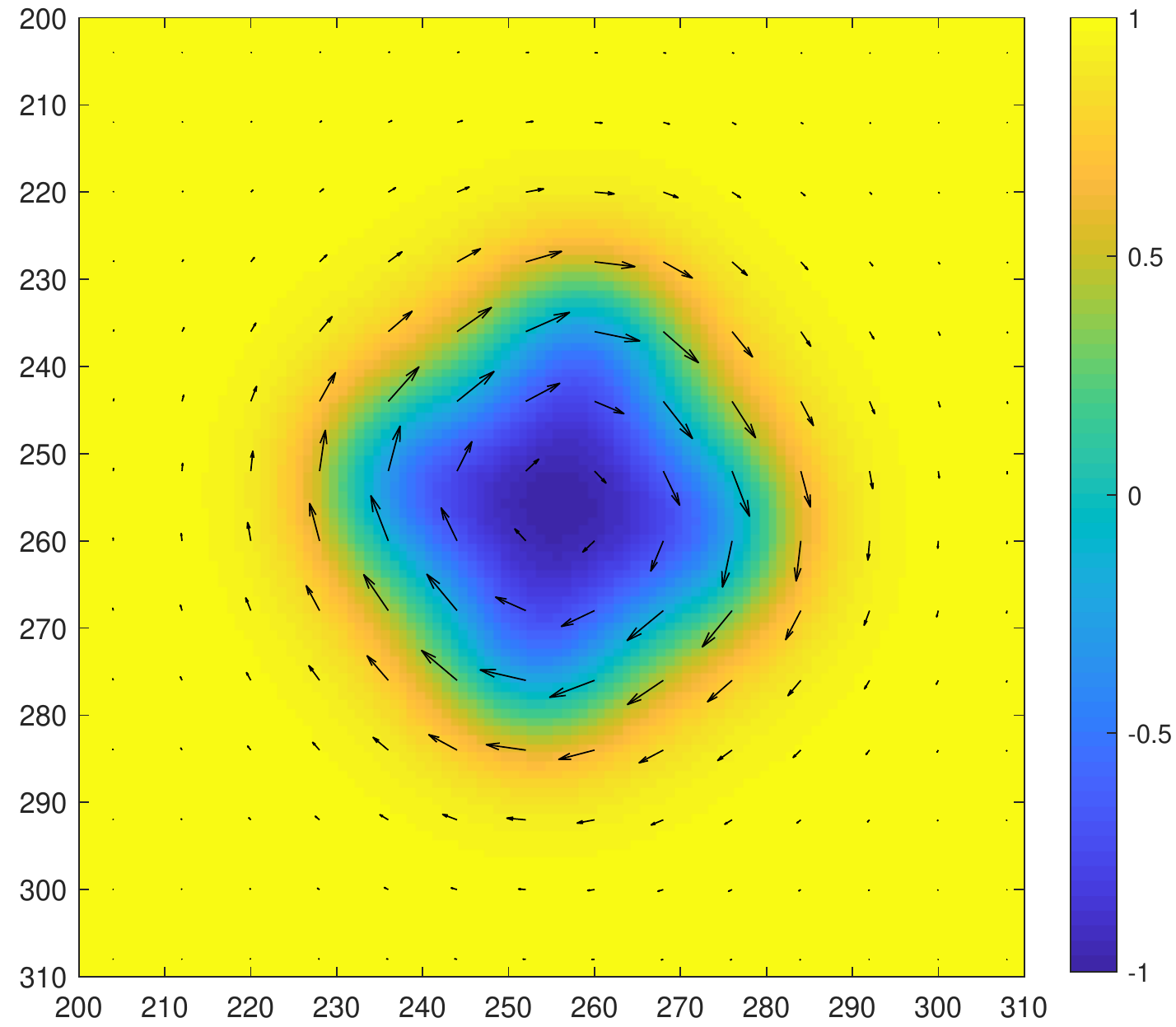}}\hfill
\subfloat[$J=3,B/J=0.024$]{\includegraphics[width=0.49\textwidth]{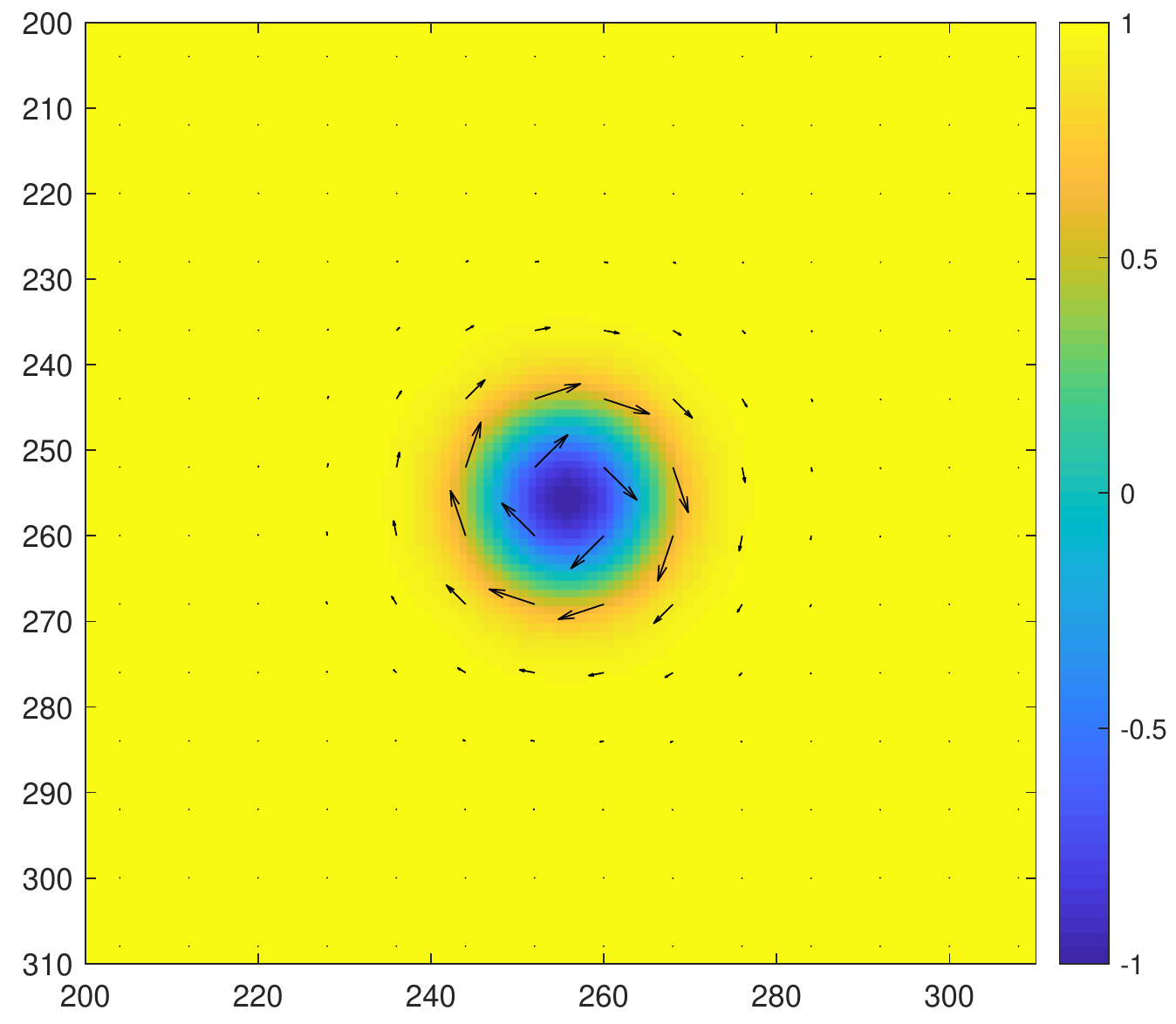}}\vfill
\subfloat[$J=1,B/J=0.018$]{\includegraphics[width=0.49\textwidth]{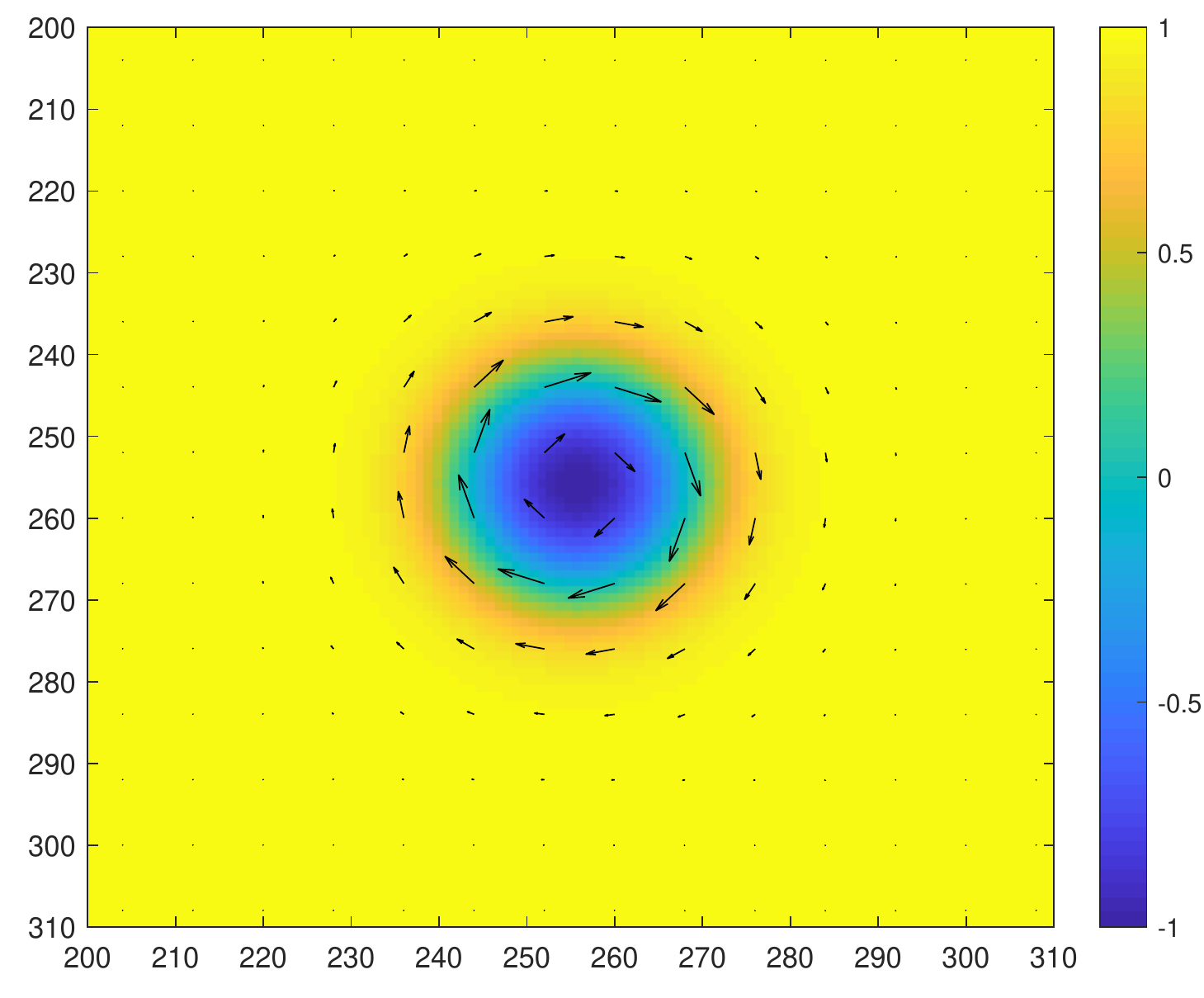}}\hfill
\subfloat[$J=5,B/J=0.018$]{\includegraphics[width=0.49\textwidth]{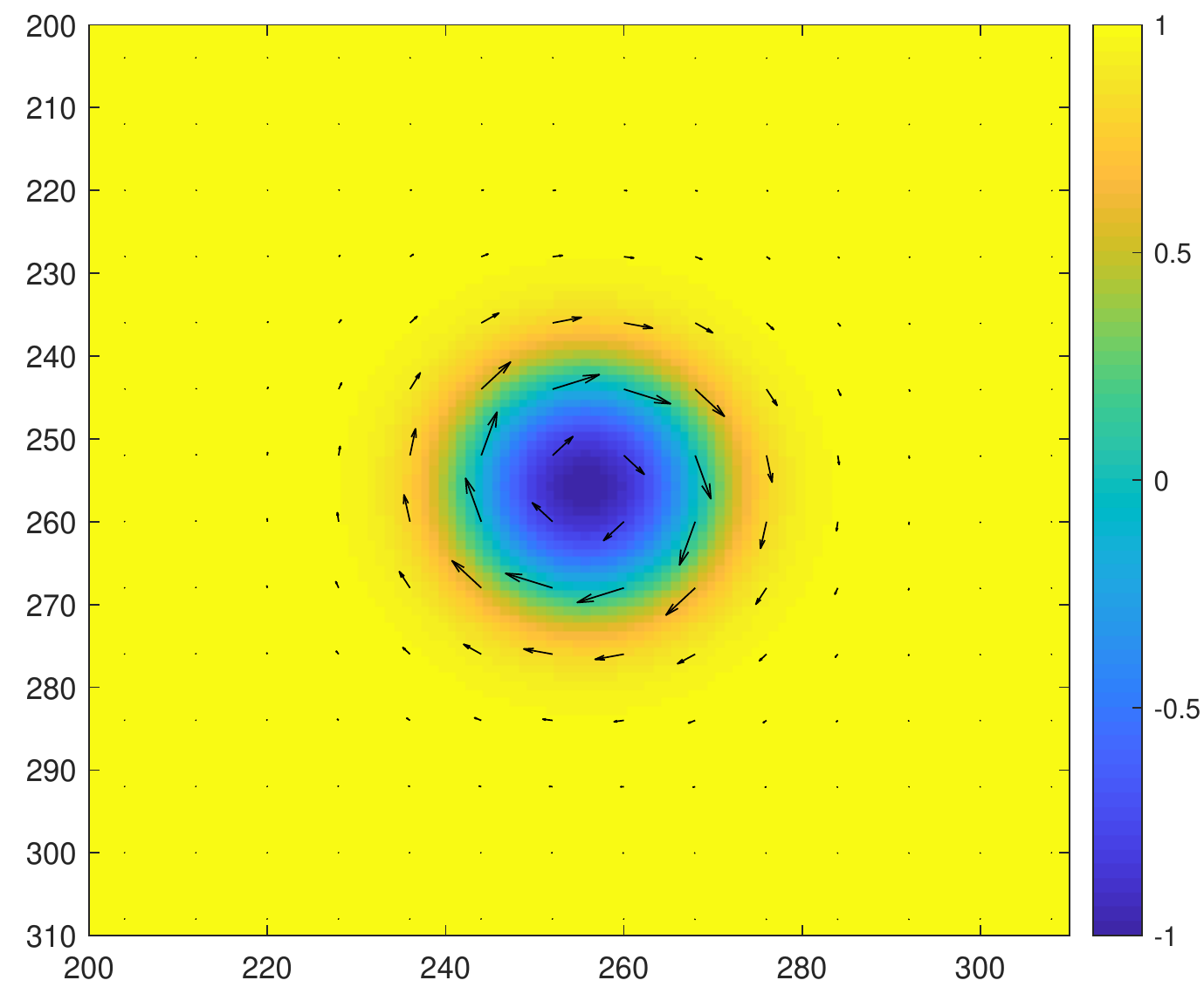}}\vfill
\caption{Simulation result with an isolated skyrmion. The heat map represents the magnitude of $n_z$ and each  arrow  represents $(n_x,n_y)$. One can find from  the value of $J$ has little effect when using dimensionless parameters.}
\label{fig:simexamplessingle}
\end{figure}

\begin{figure}
\subfloat[$J=3,B/J=0.012$]{\includegraphics[width=0.49\textwidth]{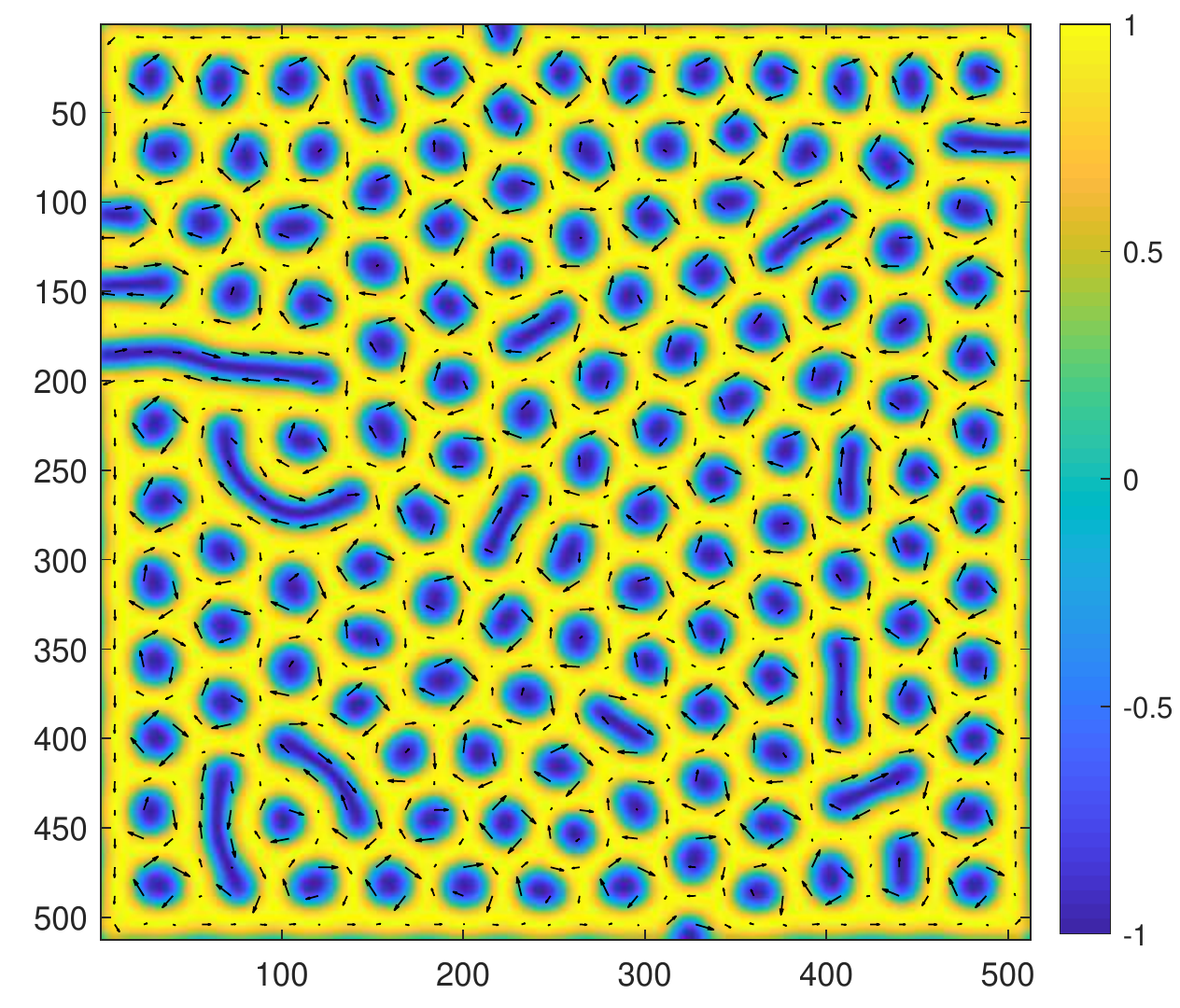}}\hfill
\subfloat[$J=3,B/J=0.024$]{\includegraphics[width=0.49\textwidth]{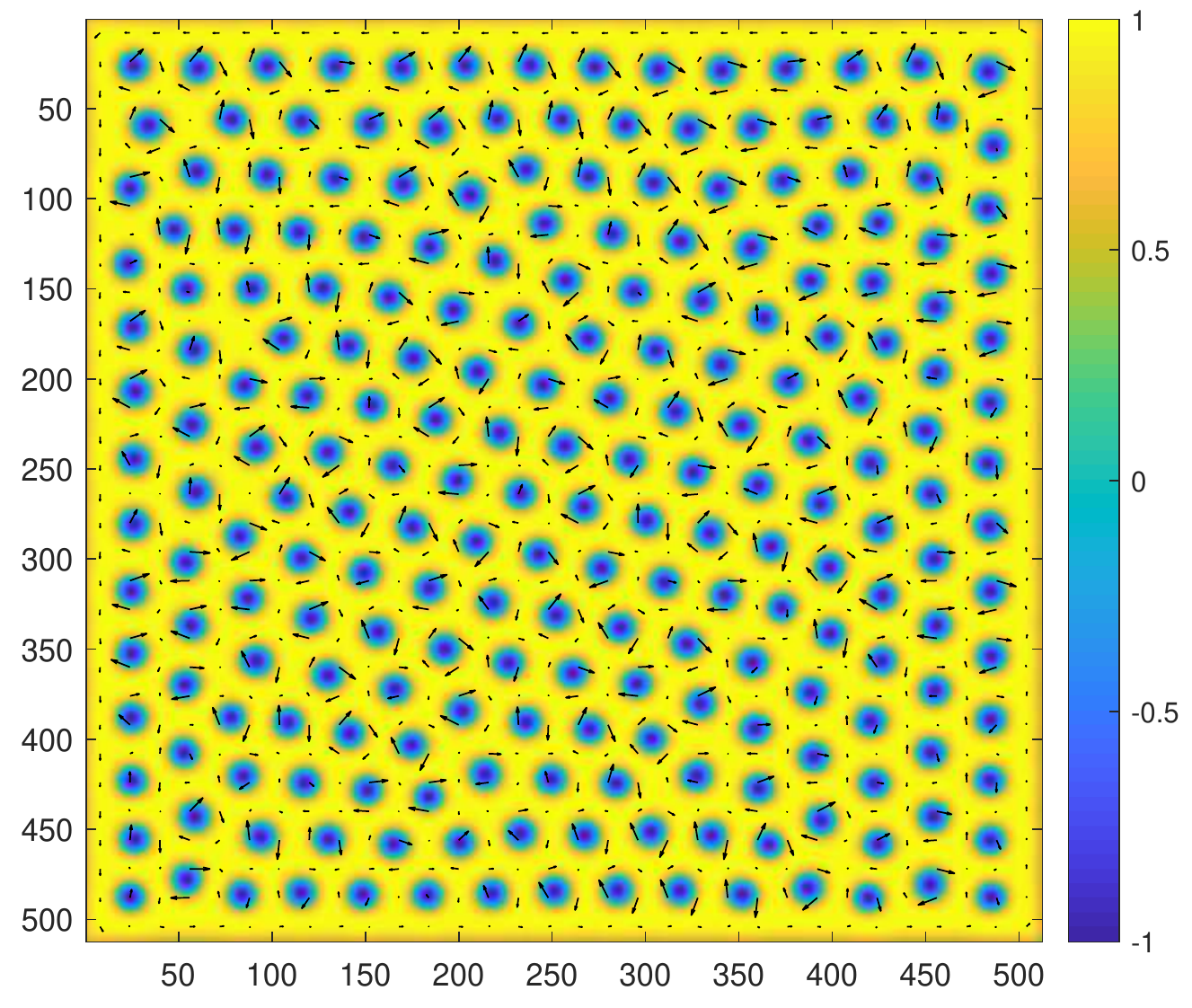}}\vfill
\subfloat[$J=1,B/J=0.018$]{\includegraphics[width=0.49\textwidth]{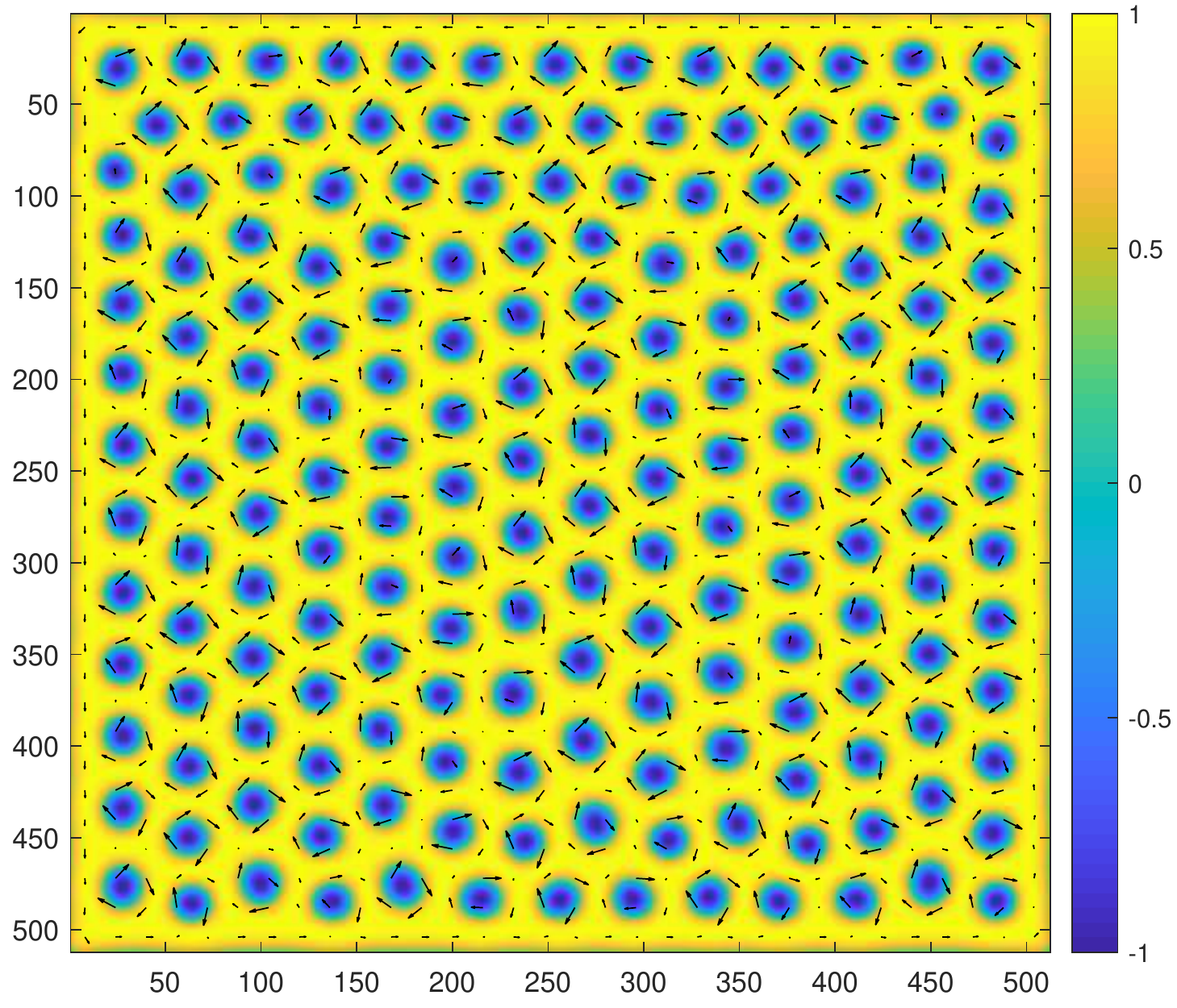}}\hfill
\subfloat[$J=5,B/J=0.018$]{\includegraphics[width=0.49\textwidth]{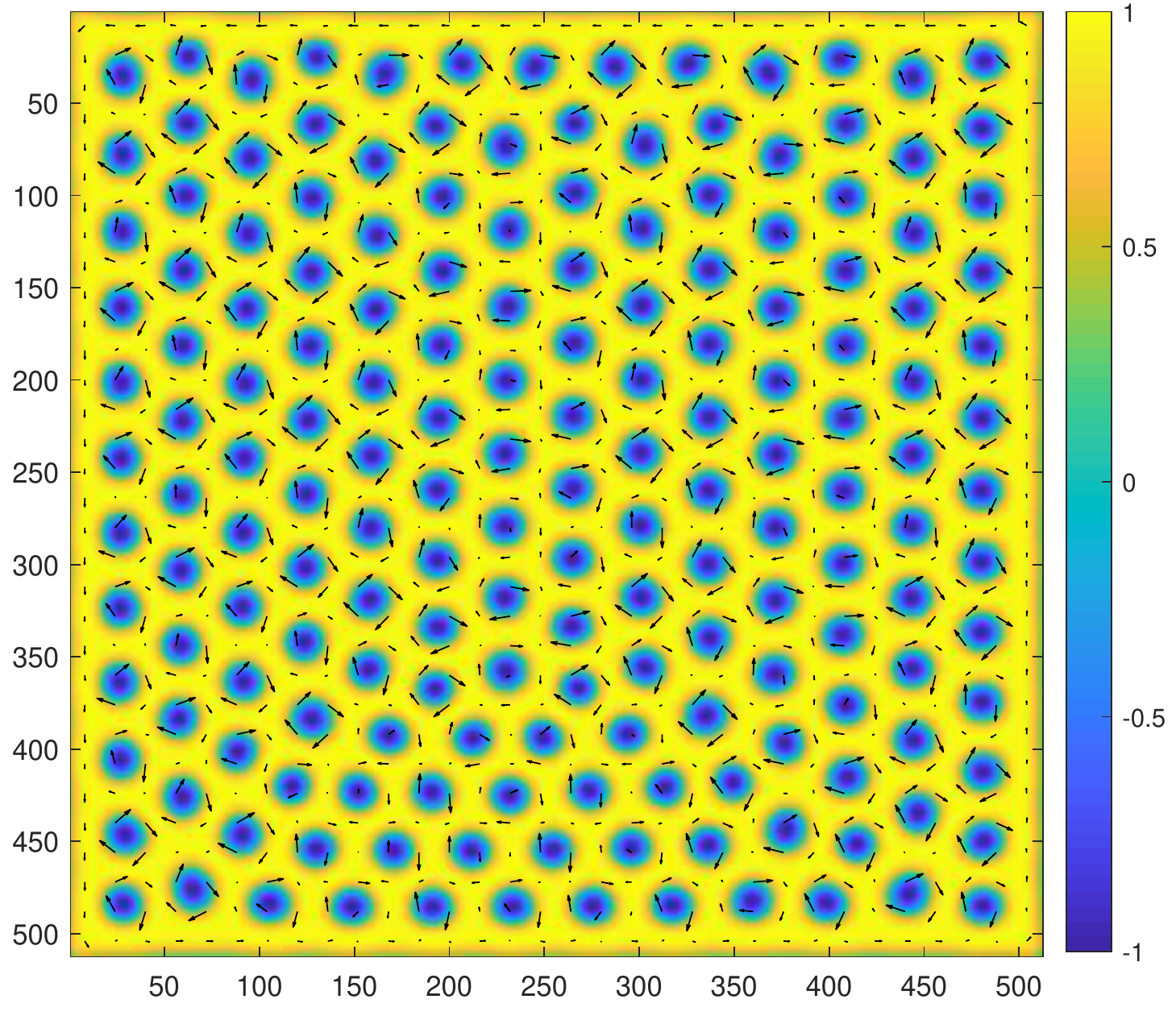}}\vfill
\caption{Simulation result  with a skyrmion phase. One can notice that the value of $J$ has little effect when using dimensionless parameters.}
\label{fig:simexamples}
\end{figure}

In the lattice simultion, we use the initial condition and the parameters in Sec.~\ref{sec:3}, and we  obtain isolated skyrmions as well as skyrmion phases. Some examples are shown in Figs.~\ref{fig:simexamplessingle} and \ref{fig:simexamples}.

\begin{figure}
\centering
\subfloat[We first calculate the isoheight contours with $n_z=0.5$ (dashed lines).]{\includegraphics[width=0.33\textwidth]{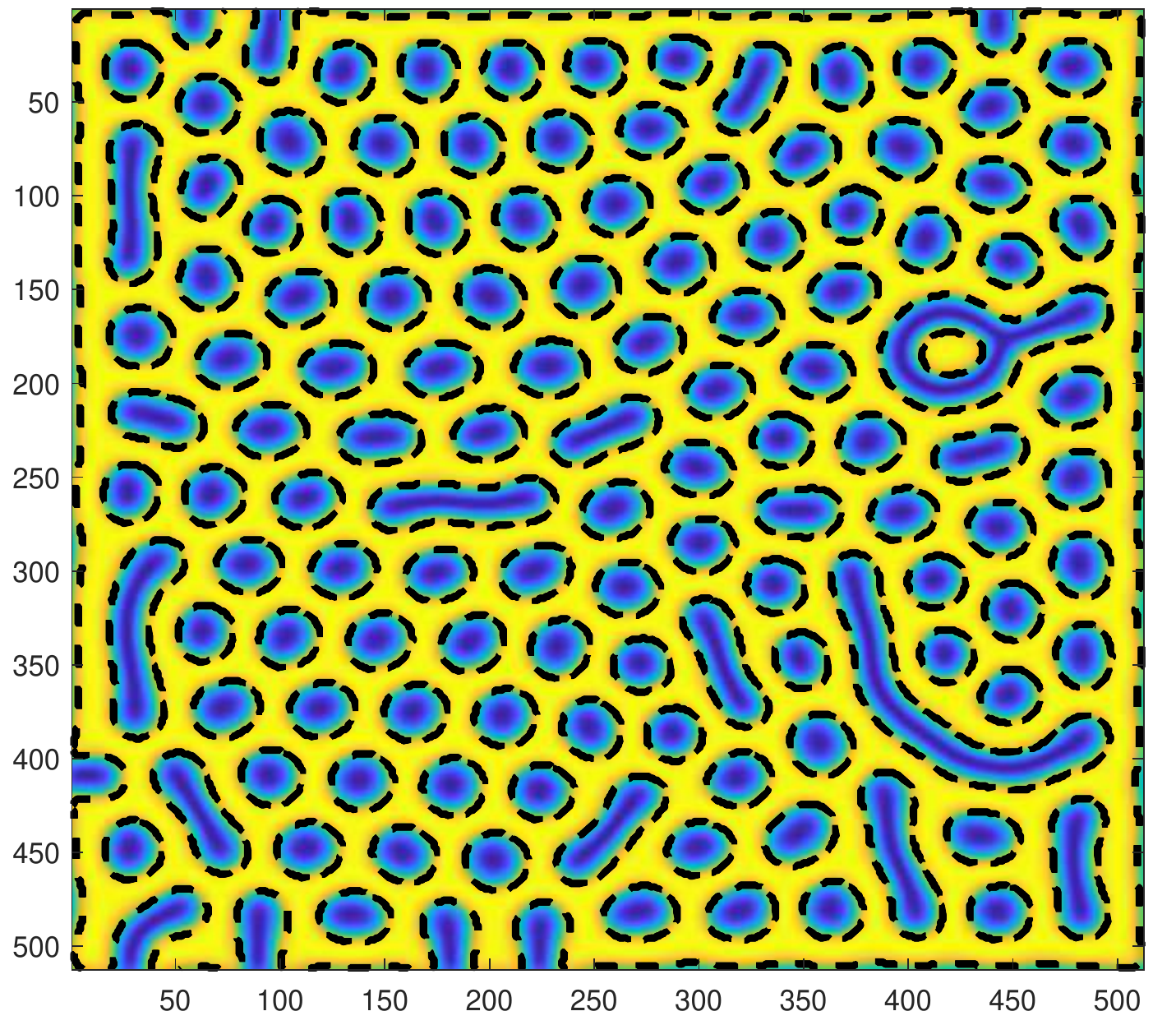}}
\hfill
\subfloat[Then we remove   contours adjacent to the edge.]{\includegraphics[width=0.33\textwidth]{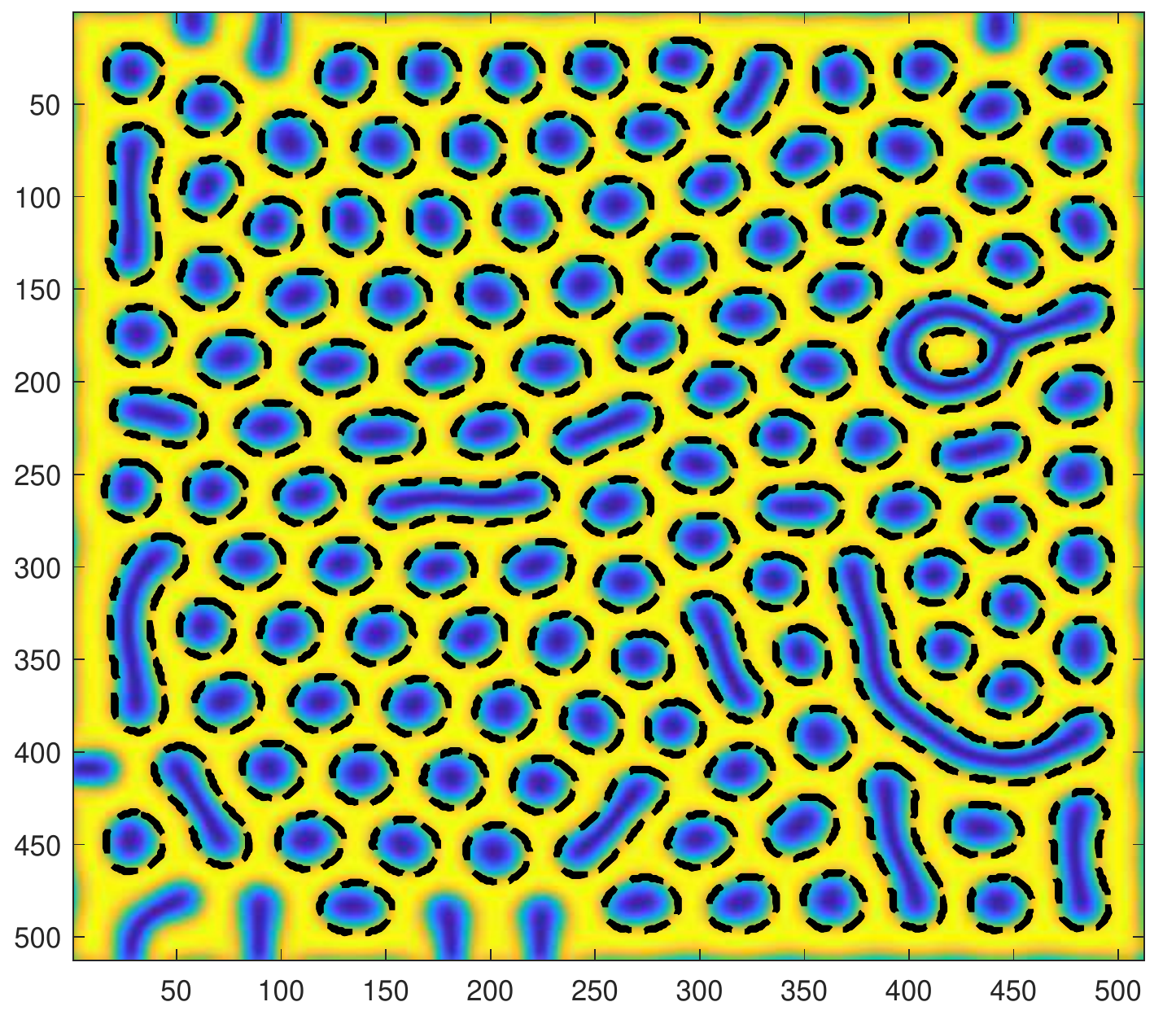}}
\hfill
\subfloat[Then we remove the contours with radius larger than $150\%$ of the median radius of the contours.]
{\includegraphics[width=0.33\textwidth]{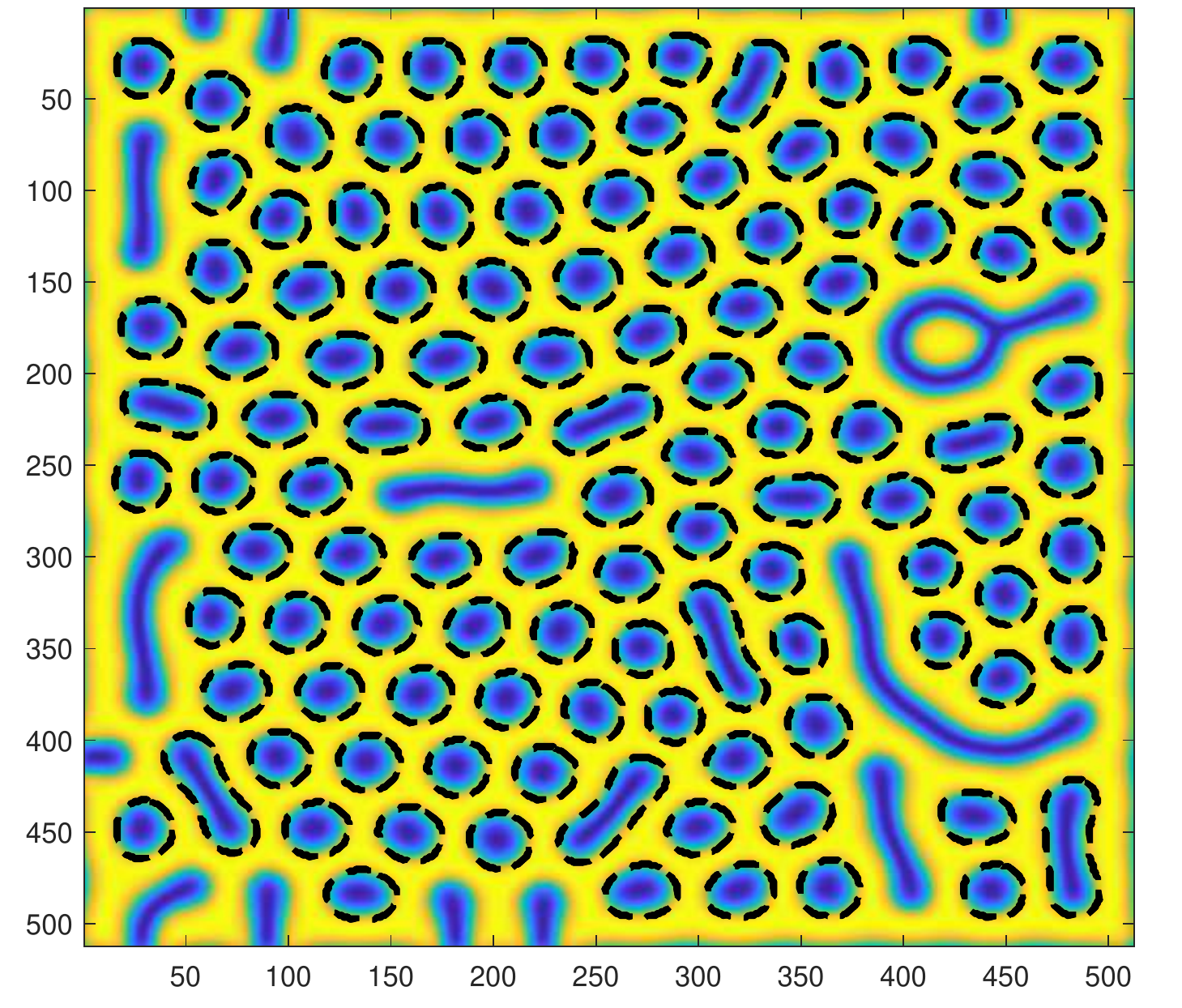} }\vfill
\caption{Illustration on how we calculate the radii of the skyrmions. $J=1,B/J=0.012$. There are $115$ skyrmions with average radius $r_s\approx 15.5176$.}
\label{fig:calcradii}
\end{figure}

To approximately calculate the skyrmion radius $r_s$, we use the isoheight contour of $n_z=0.5$, i.e. we count the number $N$ of sites with $n_z<0.5$, and the radius is estimated as  $r_s=\sqrt{N/\pi}$. For an isolated skyrmion, this procedure is straightforward. For a skyrmion phase, we first calculate the isoheight contours of $n_z=0.5$, then we discard the contours adjacent to the edge. After that, we remove those contours with radii larger than $150\%$ of the median radius. As an example, we establish the case for $J=1,B/J=0.12$ in Fig.~\ref{fig:calcradii}. In Fig.~\ref{fig:calcradii}.(c), we obtain $115$ skyrmions with  average radius $r_s\approx 15.5176$.

\begin{figure}
\includegraphics[width=0.7\textwidth]{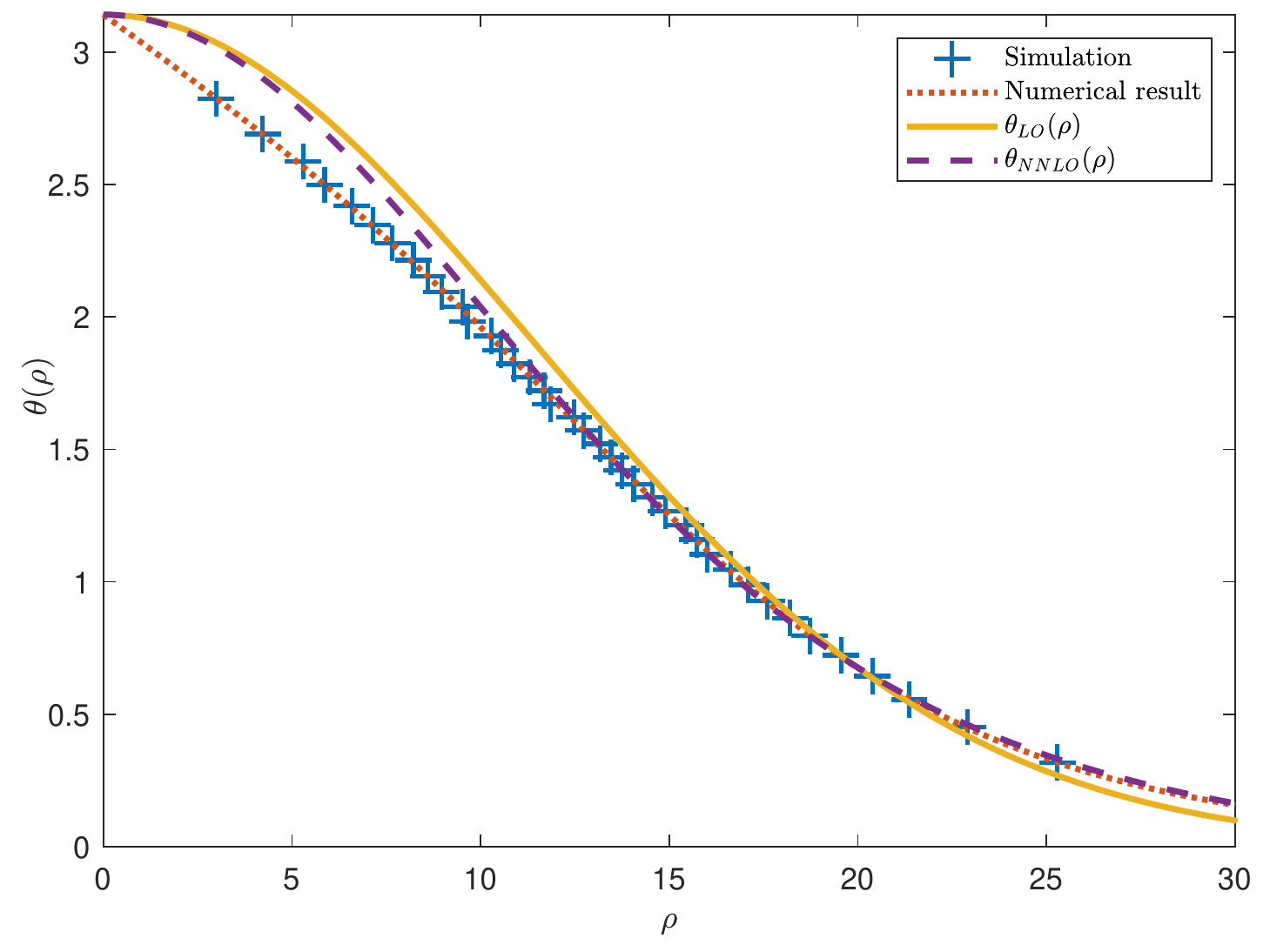}
\caption{Comparison of the simulation result for an isolated skyrmion, the numerical result, $\theta_{\rm LO}$ and $\theta_{\rm NNLO}$ with $J=1$, $D=0.18$ and $B=0.018$. The plus signs represent the average radii of the isoheight contours of $n_z$ in simulation. The dotted line represents the numerical result, the solid line represents $\theta_{\rm LO}$, and the dashed line represents $\theta_{\rm NNLO}$. The numerical result is very close to the simulation result.}
\label{fig:simulationThetaRho}
\end{figure}

In Sec.~\ref{sec:2} and Sec.~\ref{sec:3}, we find that both the numerical results and simulation results can by well fitted by our analytical calculation results. In fact, the numerical results are very closed to the results of isolated skyrmions in the simulation, it is therefore sufficient to verify only the numerical results in the case of isolated skyrmions. We show an example for $J=1$ and $B/J=0.018$~(same as Fig.~\ref{fig:simexamplessingle}.(c)) in Fig.~\ref{fig:simulationThetaRho}.

\begin{thebibliography}{99}
\bibliographystyle{unsrt}

\bibitem{nagaosa}
N. Nagaosa and Y. Tokura, Nat. Nanotechnol. \textbf{8}, 899 - 911 (2013).

\bibitem{Fert}
A. Fert, N. Reyren and V. Cros, Nat. Rev. Mater. \textbf{2}, 17031 (2017).

\bibitem{Muhlbauer}
S. M\"{u}hlbauer et. al. Science \textbf{323}, 915 - 919 (2009).

\bibitem{Yu1}
X. Z. Yu et. al. Nature \textbf{465}, 901 - 904 (2010).

\bibitem{2DSkyrmion2} 
S. Heinze et. al. Nat. Phys. \textbf{7}, 713 - 718 (2011);

C. Pfleiderer, Nature Phys. \textbf{7}, 673 - 674 (2011).

\bibitem{ultralow}
F. Jonietz et. al. Science \textbf{330}, 1648 - 1651 (2010), arXiv:1012.3496;

X. Z. Yu et. al. Nat. Commun. \textbf{3}, 988 (2012).

\bibitem{NParameterAndLinear}
A. Bogdanov and A. Hubert, J. Magn. Magn. Mater. {\bf 138} 255 (1994);

A. Bogdanov and A. Hubert, Phys. Stat. Sol. b  186, 527  (1994);

A. Bogdanov and A. Hubert, J. Magn. Magn. Mater. {\bf 195} 182 (1999).

\bibitem{NParameterArcTan} 
A. O. Leonov {\it et al.}, New J. Phys. 18, 06500 (2016).

\bibitem{Thiele}
A. Thiele, Phys. Rev. Lett. \textbf{30}, 230 - 233 (1973).

\bibitem{ThieleStudy}
K. Everschor et. al. Phys. Rev. B \textbf{86}, 054432 (2012), arXiv:1204.5051;

\bibitem{ThieleStudyAndJDB}
J. Iwasaki, M. Mochizuki, and N. Nagaosa, Nat. Nanotechnol. \textbf{8}, 742 - 747, (2013),  arXiv:1310.1655;

J. Iwasaki, M. Mochizuki, and N. Nagaosa, Nature Commun. \textbf{4}, 1463, (2013).

\bibitem{Jiang} 
W. Jiang et. al. Nat. Phys. \textbf{13}, 162 - 169 (2016), arXiv:1603.07393.

\bibitem{bilayer1}
X. Zhang et. al. Nat. Commun. \textbf{7}, 10293 (2016), arXiv:1504.02252;

\bibitem{bilayer2}
W. Koshibae and N. Nagaosa, Sci. Rep. \textbf{7}, 42645 (2017).

\bibitem{Zhang2}
X. Zhang et. al. Sci. Rep. \textbf{6}, 24795 (2016), arXiv:1504.01198.

\bibitem{pin}
Y.-H. Liu and Y.-Q. Li, J. Phys. Condens. Matter {\bf 25} 076005, arXiv:1206.5661.

\bibitem{thetarho} 
W. Jiang et. al. Phys. Reports, {\bf 704}, 1 - 49 (2017), arXiv:1706.08295.

\bibitem{arctan}
R. Nepal, U. G\"{u}ng\"{o}rd\"{u} and A. A. Kovalev, Appl. Phys. Lett. {\bf 112}, 112404 (2018), arXiv:1711.03041.

\bibitem{FreeEnergy}
J. H. Han et. al. Phys. Rev. B {\bf 82}, 094429 (2010).

\bibitem{RRMethod} 
J. K. L. MacDonald, Phys. Rev. {\bf 43} 830 (1933).

\bibitem{JDBRef1} 
M. Mochizuki, Phys. Rev. Lett. 108, 017601 (2012), arXiv:1111.5667.

\bibitem{JDBRef2} 
L. Kong and J. Zang, Phys. Rev. Lett. {\bf 111} 067203 (2013), arXiv:1308.2343.

\bibitem{alpha3}
C. Sch\"{u}tte et. al. Phys. Rev. B {\bf 90} 174434 (2014). 

\bibitem{Tchoe} 
Y. Tchoe and J. H. Han, Phys. Rev. B. {\bf 85}, 174416 (2012), arXiv:1203.0638.

\bibitem{LLG}
G. Tatara, H. Kohno and J. Shibata, Phys. Rep. {\bf 468}, 213 - 301 (2008), arXiv:0807.2894;

\bibitem{LLG2}
J. Zang et. al. Phys. Rev. Lett. {\bf 107} 136804 (2011). 

\bibitem{program}
Ji-Chong Yang, Qing-Qing Mao  and Yu Shi, Mod. Phys. Lett. B {\bf 33}, 1950040 (2019), arXiv:1809.07149.

\bibitem{Romming}
N. Romming et. al. Science \textbf{341}, 636 - 639 (2013).

\bibitem{antiskyrmion}
S. Huang et. al. Phys. Rev. B {\bf 96} 144412 (2017); 

\bibitem{pade}
S. Tokarzewski, Journal of Computational and Applied Mathematics, {\bf 75}, 2, 259 - 280 (1996).

\bibitem{padeapp}
D. J. Broadhurst, J. Fleischer and O. V. Tarasov, Z. Phys. C{\bf 60} 287 - 302  (1993), hep-ph/9304303.




\end{thebibliography}
\end{document}